\documentclass[%
    reprint,
    superscriptaddress,
    amsmath,
    amssymb,
    aps
]{revtex4-2}

\usepackage{graphicx}
\usepackage{dcolumn}
\usepackage{bm}
\usepackage{braket}
\usepackage{siunitx}
\usepackage[dvipsnames]{xcolor}
\usepackage[colorlinks]{hyperref}
\usepackage{placeins}
\usepackage{gensymb}
\hypersetup{%
    plainpages=true,
    breaklinks=true,
    hypertexnames=false,
    pageanchor=true,
    colorlinks=true,
    linkcolor={green},
    citecolor={blue},
    urlcolor={blue},
    anchorcolor={black}
}

\newcommand{\fref}[1]{Fig.~\ref{#1}}

\newcommand{\appref}[1]{\ref{#1}}

\newcommand{\matel}[1]{|\langle 0 | \hat{#1} | 1 \rangle|}

\begin{document}

\preprint{APS/123-QED}

\title{Temperature and Magnetic-Field Dependence of Energy Relaxation\\in a Fluxonium Qubit}

\def\affilPhysics{Department of Physics, Massachusetts Institute of Technology, Cambridge, MA 02139, USA}
\def\affilLL{Lincoln Laboratory, Massachusetts Institute of Technology, Lexington, MA 02421, USA}
\def\affilRLE{Research Laboratory of Electronics, Massachusetts Institute of Technology, Cambridge, MA 02139, USA}
\def\affilPhysicsHarvard{Department of Physics, Harvard University, Cambridge, MA 02139, USA}
\def\affilAQ{\textit{Atlantic Quantum, Cambridge, MA 02139}}
\def\affilG{\textit{Google Quantum AI, Santa Barbara, CA 93111}}
\def\affilEECS{Department of Electrical Engineering and Computer Science, Massachusetts Institute of Technology, Cambridge, MA 02139, USA}
\def\affilUVic{Department of Physics \& Astronomy, University of Victoria, British Columbia, Canada}

\author{Lamia Ateshian}%
  \email{ateshian@mit.edu}
  \affiliation{\affilRLE}
  \affiliation{\affilEECS}

\author{Max~Hays}%
  \affiliation{\affilRLE}

\author{David~A.~Rower}%
  \altaffiliation[Present address: ]{\affilG}
  \affiliation{\affilRLE}
  \affiliation{\affilPhysics}

\author{Helin~Zhang}%
  \affiliation{\affilRLE}

\author{Kate~Azar}%
   \affiliation{\affilRLE}
   \affiliation{\affilEECS}
   \affiliation{\affilLL}

\author{R\'{e}ouven~Assouly}%
  \affiliation{\affilRLE}

\author{Leon~Ding}%
   \altaffiliation[Present address: ]{\affilAQ}%
   \affiliation{\affilPhysics}

\author{Michael~Gingras}%
\author{Hannah~Stickler}%
\author{Bethany~M.~Niedzielski}%
\author{Mollie~E.~Schwartz}%
  \affiliation{\affilLL}
  
\author{Terry~P.~Orlando}%
  \affiliation{\affilRLE}
  \affiliation{\affilEECS}
  
\author{Joel~\^I-j.~Wang}%
\author{Simon~Gustavsson}%
  \altaffiliation[Present address: ]{\affilAQ}
\author{Jeffrey~A.~Grover}%
  \affiliation{\affilRLE}

\author{Kyle~Serniak}%
  \affiliation{\affilRLE}
  \affiliation{\affilLL}
  
\author{William~D.~Oliver}%
  \email{william.oliver@mit.edu}
  \affiliation{\affilRLE}
  \affiliation{\affilEECS}
  \affiliation{\affilPhysics}

\date{\today}

% Abstract
\begin{abstract}
Noise from material defects at device interfaces is known to limit the coherence of superconducting circuits, yet our understanding of the defect origins and noise mechanisms remains incomplete. Here we investigate the temperature and in-plane magnetic-field dependence of energy relaxation in a low-frequency fluxonium qubit, where the sensitivity to flux noise and charge noise arising from dielectric loss can be tuned by applied flux. We observe an approximately linear scaling of flux noise with temperature $T$ and a power-law dependence of dielectric loss $T^3$ up to 100 mK. Additionally, we find that the dielectric-loss-limited $T_1$ decreases with weak in-plane magnetic fields, suggesting a potential magnetic-field response of the underlying charge-coupled defects. We implement a multi-level decoherence model in our analysis, motivated by the widely tunable matrix elements and transition energies approaching the thermal energy scale in our system. These findings offer insight for fluxonium coherence modeling and should inform microscopic theories of intrinsic noise in superconducting circuits.
\end{abstract}

\maketitle

% ---------------------------------------------------------------------------------------------------------
% Figure 1: Experimental setup
% ---------------------------------------------------------------------------------------------------------
\begin{figure*}[ht!]
    \centering
    \includegraphics[width=\textwidth]{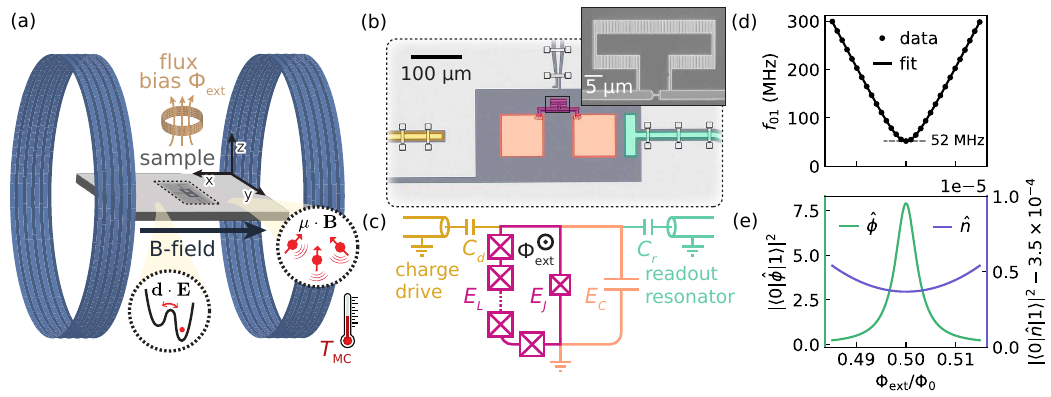}
    \caption{
    \textbf{Experimental setup and device.}
    (a) Schematic of the fluxonium sample in a superconducting magnet, which provides in-plane fields of $B \leq 100~\mathrm{G}$. 
    A coil above the chip supplies an out-of-plane flux bias $\Phi_\text{ext}$ to tune the qubit frequency. 
    Material defects---believed responsible for limiting coherence---may reside within the Josephson-junction oxides or device interfaces and interact with the qubit via electric- (charge TLS) or magnetic-dipole (surface spin) coupling. Cartoon thermometer indicates that MC temperature is adjusted for temperature-dependence measurements.
    (b) False-colored micrograph of a fluxonium circuit identical in design to the measured device. Inset: SEM image of the Josephson junction array implementing the linear inductance. 
    (c) Circuit model including readout and charge drive lines. The flux line (uncolored in micrograph) was not used in this experiment.
    (d) Low-frequency qubit spectroscopy around the degeneracy point, $|\Phi_\text{ext} - \Phi_0/2| \equiv|\delta\Phi_\text{ext}| \leq 15~\mathrm{m\Phi_0}$. The minimum qubit frequency is $f_{01} = 52~\mathrm{MHz}$, enabling energy (free-evolution) decay to probe noise down to this frequency. 
    (e) Squared magnitude of phase ($\matel{\phi}^2$) and Cooper-pair number ($\matel{n}^2$) matrix elements around half-flux. Maximum of $\matel{\phi}^2$ leads to maximal sensitivity of phase-coupled noise at $\Phi_\text{ext} / \Phi_0 = 0.5$.
    }
    \label{fig:setup}
\end{figure*}

% ---------------------------------------------------------------------------------------------------------
% Introduction
% ---------------------------------------------------------------------------------------------------------
\section{Introduction}\label{sec:intro}
Decoherence arising from the coupling of quantum systems to uncontrolled environmental degrees of freedom poses a significant challenge for solid-state quantum computing platforms, including superconducting qubits
~\cite{deleon2021materials,oliver2013materials,kjaergaardSuperconductingQubitsCurrent2020, gambettaBuildingLogicalQubits2017, siddiqiEngineeringHighcoherenceSuperconducting2021}. 
In particular, material defects located at device surfaces and interfaces provide relaxation and dephasing channels that can limit coherence
~\cite{yanRotatingframeRelaxationNoise2013, yanFluxQubitRevisited2016, yanSpectroscopyLowfrequencyNoise2012, bylanderNoiseSpectroscopyDynamical2011, sendelbachMagnetismSQUIDsMillikelvin2008,quintanaObservationClassicalQuantumCrossover2017,yoshiharaDecoherenceFluxQubits2006, bialczakFluxNoiseJosephson2007, kakuyanagiDephasingSuperconductingFlux2007, kochChargeInsensitiveQubit2007,martinisDecoherenceJosephsonQubits2005, lisenfeldElectricFieldSpectroscopy2019, crowley2023disentangling}, degrading the fidelity of quantum operations. 
As a primary example, it has long been observed that ensembles of two-level system (TLS) defects -- tunneling systems that reside in the disordered dielectric regions of devices -- couple parasitically to quantum circuits via their electric dipole moments, resulting in both broadband charge noise and individual dropouts in energy lifetimes~\cite{Phillips1987TwoLevelStates, mullerUnderstandingTwolevelsystemsAmorphous2019,paladinoNoiseImplicationsSolidstate2014,lisenfeldElectricFieldSpectroscopy2019,martinisDecoherenceJosephsonQubits2005}. 
A potentially distinct family of defects comprise those with intrinsic magnetic dipole moments, which generally constitute a dominant source of low-frequency $1/f$ magnetic flux noise~\cite{laforestFluxvectorModelSpin2015,sendelbachMagnetismSQUIDsMillikelvin2008,degraafDirectIdentificationDilute2017,lantingEvidenceTemperaturedependentSpin2014, lantingProbingEnvironmentalSpin2020,braumullerCharacterizingOptimizingQubit2020}, often limiting dephasing times in flux-tunable qubits
~\cite{yoshiharaDecoherenceFluxQubits2006, bialczakFluxNoiseJosephson2007, kakuyanagiDephasingSuperconductingFlux2007, kochChargeInsensitiveQubit2007, bylanderNoiseSpectroscopyDynamical2011}.
Such surface spins~\cite{laforestFluxvectorModelSpin2015, bluvsteinExtendingQuantumCoherence2019, joosProtectingQubitCoherence2021} are suspected to consist of surface adsorbates such as molecular oxygen and hydrogen~\cite{degraafDirectIdentificationDilute2017, degraafSuppressionLowfrequencyCharge2018, Jayaraman2024} and may exhibit correlations or clustering behavior that impact their dynamics~\cite{laforestFluxvectorModelSpin2015, deFluxNoiseLow2019, antonMagneticFluxNoise2013}. 
Notably, in qubits operated at frequencies $f_{01} \lesssim 1~\mathrm{GHz}$, such as the flux qubit, fluxonium, or fluxmon, $1/f$ noise may also result in energy relaxation~\cite{yanFluxQubitRevisited2016, Zhang2021Universal, quintanaObservationClassicalQuantumCrossover2017, smith2020superconducting, hays2025non}.

Many efforts have investigated the physical and chemical properties of material defects in quantum circuits by probing their responses to quantities such as DC electric field~\cite{lisenfeldElectricFieldSpectroscopy2019}, magnetic field~\cite{degraafDirectIdentificationDilute2017, rower2023evolution, Jayaraman2024, gunzler2025spin}, microwave power~\cite{martinisDecoherenceJosephsonQubits2005, gao2007noise, Burnett2016Analysis, crowley2023disentangling}, temperature~\cite{burnettEvidenceInteractingTwolevel2014,crowley2023disentangling,antonMagneticFluxNoise2013,quintanaObservationClassicalQuantumCrossover2017}, circuit geometry~\cite{woodsDeterminingInterfaceDielectric2019, yanFluxQubitRevisited2016, kumarOriginReductionMagnetic2016, braumullerCharacterizingOptimizingQubit2020}, and chemical treatments~\cite{degraafDirectIdentificationDilute2017, degraafSuppressionLowfrequencyCharge2018, kumarOriginReductionMagnetic2016, mergenthalerEffectsSurfaceTreatments2021, Jayaraman2024}.
However, despite these intense investigations into understanding and mitigating the effects of charge TLS and surface spins~\cite{degraafChemicalStructuralIdentification2022, paladinoNoiseImplicationsSolidstate2014, mullerUnderstandingTwolevelsystemsAmorphous2019, degraafChemicalStructuralIdentification2022, davis_1f_2018, navaaquinoFluxNoiseDisordered2022}, their exact origins have remained elusive.
Additionally, while there is some evidence that the charge- and flux-noise defect dynamics are correlated~\cite{degraafDirectIdentificationDilute2017, degraafSuppressionLowfrequencyCharge2018}, the mechanisms of such correlations are unknown.
Addressing decoherence from material defects is crucial for building a large-scale superconducting quantum computer~\cite{google2023suppressing}, motivating continued efforts to investigate these intrinsic noise sources. 

Qubits themselves are spectrometers that can be operated as probes of their surrounding noise environments~\cite{yanRotatingframeRelaxationNoise2013, yanSpectroscopyLowfrequencyNoise2012, yanFluxQubitRevisited2016, bylanderNoiseSpectroscopyDynamical2011, Schoelkopf2002Spectrometers}. 
We focus on the fluxonium qubit~\cite{Manucharyan2009, nguyenHighCoherenceFluxoniumQubit2019, earnest2018realization}, which features high coherence~\cite{Somoroff2023, Nguyen2019, Pop2014Coherent} and has been used to implement high-fidelity single-~\cite{Rower2024CounterRotating, somoroff2024fluxonium, Zhang2021Universal}
and two-qubit~\cite{Ding2023HighFidelity, dogan2023two, moskalenko2022high} gates.
With transition frequencies as low as several MHz~\cite{najera2024high, Zhang2021Universal} as well as widely tunable matrix elements, the
fluxonium serves as a natural probe of charge- and flux-coupled low-frequency noise.
% , understanding its noise mechanisms is of especially high interest.
Several recent experiments have investigated the flux-bias dependence of fluxonium energy relaxation~\cite{Pop2014Coherent, nguyenHighCoherenceFluxoniumQubit2019, Zhang2021Universal, Sun2023FluxoniumLossMech, Ardati2024Bifluxon, Hazard2019Nanowire, spiecker2023szilard}. 
However, some commonly observed features are difficult to explain with circuit-level loss models, such as the rich structure produced by coupling to a discrete environment of TLS~\cite{spiecker2023szilard,Pop2014Coherent, Sun2023FluxoniumLossMech}.
Furthermore, studies of fluxonium coherence to date have focused primarily on its frequency (or, equivalently, flux-bias) dependence. 

In this work, we use a fluxonium qubit to probe the frequency, temperature, and magnetic-field dependence of low-frequency flux and charge noise, revealing trends that offer empirical constraints on candidate noise mechanisms in superconducting qubits.
We designed our device to have a minimum frequency of 52 MHz at the degeneracy point, enabling us to probe low-frequency $1/f$ noise directly with $T_1$ measurements. 
We compare our data to a multi-level decoherence model, motivated by the widely tunable transition strengths and level spacings approaching the thermal energy scale $k_BT$.
We observe an increase in flux noise with temperature $T$ at frequencies $1~\mathrm{MHz} \lesssim f \lesssim 100~\mathrm{MHz}$ consistent with a linear $T$ scaling of the flux noise magnitude. 
We also find an apparent power-law scaling $T^3$ of the charge noise with temperature. We explore the possibilities that this charge-noise scaling arises from either higher-level contributions to the depolarization rate or a TLS-induced decoherence mechanism that does not saturate with temperature~\cite{Phillips1987TwoLevelStates, Behunin2016Dimensional}.
Finally, in-plane magnetic-field dependence measurements are consistent with an increase in dielectric loss with field, which may further suggest a link between charge- and phase-coupled defects~\cite{degraafSuppressionLowfrequencyCharge2018}.
Our results reveal new trends in fluxonium coherence that can serve as benchmarks for future microscopic theories of charge and flux noise.

% ---------------------------------------------------------------------------------------------------------
% Fluxonium qubit design
% ---------------------------------------------------------------------------------------------------------
\section{Fluxonium Qubit Design}\label{sec:device}
Our sample is an aluminum fluxonium qubit fabricated on a silicon substrate (Fig.~\ref{fig:setup}b). Al-AlO$_x$-Al Josephson junctions serve as both the small, main tunneling element and the larger array junctions realizing the shunt inductance~\cite{Randeria2024CQPS}. 
% 
% The junction tunnel barriers and interface oxides are known to host baths of material defects that parasitically couple to the qubit via their electric (magnetic) dipole moments, serving as a candidate source of charge (phase)-coupled noise (Fig.~\ref{fig:setup}a).
% 
For our magnetic-field study, in-plane magnetic fields of up to $B = 100~\mathrm{G}$ are applied to the device with a pair of superconducting electromagnets (Fig.~\ref{fig:setup}a), similar to those used in Ref.~\cite{rower2023evolution}. 
A small coil above the sample provides an out-of-plane flux bias $\Phi_\text{ext}$ to tune the qubit frequency.
The magnets and sample are mounted at the mixing chamber (MC) stage of a dilution refrigerator (DR), whose temperature is controlled with applied heating currents up to a value of $T_\text{MC} \approx 100~\mathrm{mK}$ (see ~\ref{app:setup} for more details on experimental setup).

The fluxonium circuit (Fig.~\ref{fig:setup}c) is modeled by the Hamiltonian
\begin{align}
    \hat{H}_\text{qb} = 4E_C\hat{n}^2 - E_J \cos\hat{\phi} + \frac{1}{2} E_L (\hat{\phi}  - \phi_\text{ext}) ^2, \label{eq:hamiltonian}
\end{align}
where $E_C/h = 0.957$ GHz, $E_J/h = 6.814$ GHz and $E_L/h = 0.560$ GHz are the charging, Josephson, and inductive energies, respectively, obtained from fits to spectroscopy measurements (see~\ref{app:qb-res-spec}). The phase bias $\phi_\text{ext} = 2\pi \Phi_\text{ext} / \Phi_0$ arises from the externally-applied magnetic flux. 
For the magnetic-field study, we focus on the spectral band between the minimum frequency $f_{01} = 52~\mathrm{MHz}$ and 300 MHz, shown in Fig.~\ref{fig:setup}d.
We choose this range for two primary reasons. First, it contains the Zeeman frequency of a free electron spin with gyromagnetic ratio $\gamma \approx 2.8$ MHz/G at the maximum applied magnetic field, enabling the possible identification of free-spin features in the relaxation spectrum~\cite{gunzler2025spin}. Second, the depolarization trends in this range are reasonably captured by circuit-level loss models, in contrast to the higher-frequency loss that displays large variation from TLS.
% 
% Most previous noise studies in this frequency range have used higher-frequency qubits, necessitating noise spectroscopy techniques that probe pure-dephasing noise such as dynamical decoupling~\cite{bylanderNoiseSpectroscopyDynamical2011} or driven relaxation such as spin-locking~\cite{yanRotatingframeRelaxationNoise2013}.
Most previous noise studies in this frequency range have used higher-frequency qubits, necessitating noise spectroscopy techniques that probe pure-dephasing noise (e.g., dynamical decoupling~\cite{bylanderNoiseSpectroscopyDynamical2011}) or driven relaxation (e.g., spin-locking~\cite{yanRotatingframeRelaxationNoise2013}).
However, the lower frequency afforded by our fluxonium design enables us to probe this frequency range directly with (free-evolution) $T_1$ measurements.

We model the depolarization rate $\Gamma_1 = 1/T_1$ from a combination of noise sources with Fermi's golden rule,
% \begin{equation}
%     \Gamma_1 = \sum_\eta \frac{\alpha_\eta^2}{\hbar^2} |\bra{0}{\hat{D}_\eta}\ket{1}|^2 S_\eta^+(\omega_{01}), \label{eq:gamma1}
% \end{equation}
\begin{equation}
    \Gamma_1 = \sum_\lambda \frac{2}{\hbar^2} |\bra{0}{\hat{D}_\lambda}\ket{1}|^2 S_\lambda(\omega_{01}), \label{eq:gamma1}
\end{equation}
where $S_\lambda(\omega) = \frac{1}{2} \int_{-\infty}^\infty \langle \lambda(0) \lambda(t) + \lambda(t)\lambda(0)\rangle e^{-i\omega t} dt$
is the symmetrized, bilateral power spectral density of the noise $\lambda(t)$,
and the coupling of each noise source to the qubit is modeled by an interaction Hamiltonian 
$\hat{H}_\text{int} = \hat{D}_\lambda \lambda(t)$ (see~\ref{app:models}).
We primarily consider noise sources that couple to the qubit charge (expressed in terms of Cooper-pair number) and phase operators (see Fig.~\ref{fig:setup}e).
The phase transition matrix element --- and thereby the qubit sensitivity to flux noise --- is maximal at the degeneracy point $\Phi_\text{ext} = \Phi_0/2$, referred to as half-flux, and decreases quickly as the bias is tuned away from the degeneracy point. 
Since tuning flux changes the relative strength of charge- and flux-noise-induced $\ket{0} \leftrightarrow \ket{1}$ transitions, it also adjusts the sensitivity to charge- and phase-coupled noise.

% ---------------------------------------------------------------------------------------------------------
% Figure 2: Flux dependence of T1
% ---------------------------------------------------------------------------------------------------------
\begin{figure}[t!]
    \centering
    \includegraphics[width=\columnwidth]{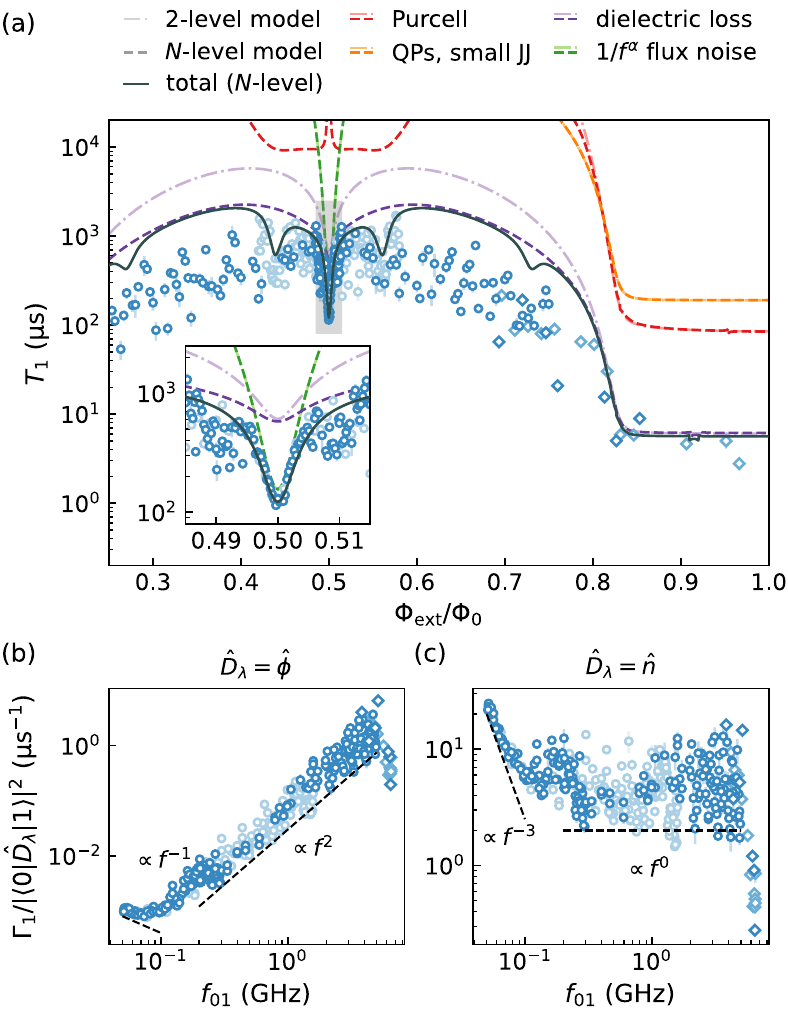}%
    \caption{
    \textbf{Baseline $T_1$ characterization and models.}
    (a) Measured $T_1$ versus external flux bias at base temperature $(T_\text{MC} \approx 35~\mathrm{mK})$ and in the absence of applied magnetic field $(B = 0~\mathrm{G})$.
    Coherence data aggregated from three cooldowns (indicated by color) show consistency. We prepare the initial state either by post-selection (indicated by circular markers) or by applying a $\pi$-pulse (indicated by diamond markers). Inset zooms in on the region around half-flux, $|\delta\Phi_\text{ext}| \leq 15~\mathrm{m\Phi_0}$, corresponding to the qubit frequency range in Fig.~\ref{fig:setup}d. 
    Light dash-dotted (dark dashed) lines indicate individual loss mechanisms modeled with two levels ($N$ levels), and the solid line indicates the total loss calculated by the $N$-level model (see main text and~\ref{app:heating}).
    (b, c) Depolarization rates versus frequency, normalized by squared magnitude of transition matrix elements (same data as panel (a)): $\Gamma_1(\omega) / |\bra{0} \hat{D}_\lambda \ket{1}|^2 \propto S_\lambda(\omega)$, with $\hat{D}_\lambda = \hat{\phi}$ (b) and $\hat{n}$ (c).
    Guides for different frequency scaling behaviors are shown for comparison.
    }
    \label{fig:t1-flux-baseline}
\end{figure}

% ---------------------------------------------------------------------------------------------------------
% T1 Characterization and Modeling
% ---------------------------------------------------------------------------------------------------------
\section{$T_1$ Characterization and Modeling}\label{sec:T1-baseline}
We begin by characterizing the flux dependence of $T_1$ at base temperature in the absence of an in-plane magnetic field, shown in Fig.~\ref{fig:t1-flux-baseline}a.
% We perform energy relaxation measurements using standard time-domain methods, with the initial state prepared either by post-selection or by applying a $\pi$-pulse to the $\ket{0} \leftrightarrow \ket{1}$ transition (see ~\ref{app:T1-measurements} for details). 
We perform energy relaxation measurements using standard time-domain methods; we initialize the qubit either by post-selecting on the outcome of a preceding measurement, or (for high-frequency transitions) by preparing the $\ket{1}$ state with a $\pi$-pulse on the $\ket{0} \leftrightarrow \ket{1}$ transition (see ~\ref{app:T1-measurements} for details).
The qubit state is read out dispersively through a resonator.
We model the data with a combination of loss mechanisms, each with a different flux-bias dependence: $1/f^\alpha$ flux noise (with $\alpha \lesssim 1$), dielectric loss, Purcell decay, and quasiparticle (QP) tunneling. 

First, we consider flux biases near the degeneracy point, where phase-coupled noise dominates. The decay rate for phase-coupled noise following Fermi's Golden Rule is
% \begin{equation}
%     \Gamma_1^\Phi = \left(\frac{2\pi E_L}{\hbar\Phi_0}\right)^2 |\langle 0|\hat{\phi}|1 \rangle |^2 S_{\Phi}^+(\omega_{01}). \label{eq:gamma1-1overf}
% \end{equation}
\begin{equation}
    \Gamma_1^\Phi = 2\left(\frac{2\pi E_L}{\hbar\Phi_0}\right)^2 |\langle 0|\hat{\phi}|1 \rangle |^2 S_{\Phi}(\omega_{01}), \label{eq:gamma1-1overf}
\end{equation}
where $S_\Phi(\omega)$ is the flux noise power spectral density, which we model as~\cite{yanFluxQubitRevisited2016, yanSpectroscopyLowfrequencyNoise2012}
% \begin{equation}
%     S_{\Phi,1/f}^+(\omega) = 2A_\Phi \left(\frac{2\pi}{\omega} \right)^\alpha, \label{eq:1/f-flux-noise}
% \end{equation}
\begin{equation}
    S_{\Phi}(\omega) = A_\Phi \left(\frac{2\pi}{\omega} \right)^\alpha. \label{eq:1/f-flux-noise}
\end{equation}
Here, $A_\Phi$ is the magnitude of the noise at 1 Hz, and $\alpha$ is the exponent characterizing the frequency dependence.
We independently estimate the flux noise spectrum parameters from spin-locking~\cite{yanRotatingframeRelaxationNoise2013} and spin-echo dephasing measurements~\cite{braumullerCharacterizingOptimizingQubit2020} (see~\ref{app:low-freq-flux-noise}), which characterize the qubit longitudinal noise at frequencies $\sim 1 - 30~\mathrm{MHz}$, close to the lowest frequency probed by $T_1$ ($f_{01} = \SI{52}{MHz}$). These measurements yield $A_\Phi = (0.25~\mathrm{\mu\Phi_0})^2$ and $\alpha = 0.62$, consistent with the measured $T_1$ values near half-flux.
These apparently anomalously low values of $A_\Phi$ and $\alpha$ are a result of allowing $\alpha$ to vary (see~\ref{app:low-freq-flux-noise}). Assuming $\alpha = 1$ for the same data yields $A_\Phi = (3.7~\mathrm{\mu \Phi_0})^2$, which is consistent with the magnitude typically observed for intrinsic flux noise~\cite{Randeria2024CQPS, Ardati2024Bifluxon, rower2023evolution, braumullerCharacterizingOptimizingQubit2020, yanFluxQubitRevisited2016, yanRotatingframeRelaxationNoise2013, yanSpectroscopyLowfrequencyNoise2012, slichter2012measurement, bylanderNoiseSpectroscopyDynamical2011}.

Next, we consider $T_1$ away from half-flux, where the sensitivity to phase-coupled noise diminishes, and charge-coupled noise can limit decoherence. The decay rate for charge-coupled noise is given by
% \begin{equation}
%     \Gamma_1^Q = \left(\frac{4E_C}{e\hbar}\right)^2 |\langle 0 | \hat{n} | 1 \rangle|^2 S_Q^+(\omega_{01}). \label{eq:gamma1-diel}
% \end{equation} 
\begin{equation}
    \Gamma_1^Q = 2\left(\frac{4E_C}{e\hbar}\right)^2 |\langle 0 | \hat{n} | 1 \rangle|^2 S_Q(\omega_{01}), \label{eq:gamma1-diel}
\end{equation} 
where $S_Q(\omega)$ is the charge noise power spectral density.
We model the charge noise arising from dielectric loss as Johnson-Nyquist voltage fluctuations in the capacitive elements of the circuit~\cite{Schoelkopf2002Spectrometers}:
% \begin{equation}
%     S_{Q,\text{diel}}^+(\omega) = \frac{e^2\hbar}{E_C} \tan\delta_C (\omega) \coth\left( \frac{\hbar\omega}{2k_BT_\text{eff}} \right), \label{eq:diel-loss-noise-spectrum}
% \end{equation}
\begin{equation}
    S_{Q}(\omega) = \frac{e^2\hbar}{2E_C} \tan\delta_C (\omega) \coth\left( \frac{\hbar\omega}{2k_BT_\text{eff}} \right), \label{eq:diel-loss-noise-spectrum}
\end{equation}
where $\tan\delta_C(\omega)$ is the dielectric loss tangent, which may possess a frequency dependence.
Following prior works~\cite{nguyenHighCoherenceFluxoniumQubit2019, Pop2014Coherent}, we model this phenomenologically as $\tan\delta_C(\omega) = \tan\delta_C^0 \cdot (\omega / \omega_\text{ref})^\epsilon$, where  $\omega_\text{ref}/2\pi = 6~\mathrm{GHz}$ is an arbitrary reference frequency~\cite{Wang2015Surface}.
Close to integer flux biases ($\Phi_\text{ext} \approx m\Phi_0$ with integer $m$), where the $\ket{0}\leftrightarrow\ket{1}$ transition occurs between wavefunctions within the same potential well, the charge transition matrix element is maximal. 
Assuming that dielectric loss limits $T_1$ around this transition, we estimate an upper bound on the loss tangent of $\tan\delta_C^0 \approx  4 \times 10^{-6}$. 
Constraining our fit with this value and the parameters fixed by the flux noise measurements, $\epsilon = 0.26$ describes our data best near the degeneracy point with an effective temperature of $T_\text{eff} = 50~\mathrm{mK}$.
We estimate this effective qubit temperature by measuring the $\ket{0},\ket{1}$ state populations, finding it elevated with respect to the mixing-chamber temperature of $T_\text{MC} \approx 35~\mathrm{mK}$ (see~\ref{app:T1-measurements} for further discussion on $T_\text{eff}$).
The values of $\tan\delta_C^0$ and $\epsilon$ are consistent with those reported in prior works~\cite{nguyenHighCoherenceFluxoniumQubit2019, Sun2023FluxoniumLossMech, Pop2014Coherent}, although the physical origin of $\epsilon$ is unclear.

To address Purcell decay through the readout resonator, we model the resonator as a $\lambda/4$ transmission line inductively coupled to a $50~\Omega$ environment (see~\ref{app:models}). 
% Environmental voltage fluctuations 
A measurement of $T_\phi^\text{E}$ at the degeneracy point places an upper bound on the resonator temperature of $T_\text{res} = 70~\mathrm{mK}$, which may be elevated by noise photons in the measurement line from higher-temperature stages of the DR. This value results in an estimate of $T_1^\text{Purcell}$ 
at $\Phi_\text{ext}/\Phi_0 = 0$ of approximately $80~\mathrm{\mu s}$, which is well above the measured $T_1 \approx 5~\mathrm{\mu s}$, consistent with coherence limited by mechanisms beyond Purcell decay through the resonator.

Tunneling of QPs across the small junction may also induce qubit-state transitions~\cite{serniak2018hot, Pop2014Coherent, Catelani2011Relaxation, harrington2024synchronous}, but to limit the depolarization rate in our device, a relative QP density of $x_\text{qp} \sim 10^{-6}$ would be required, more than an order of magnitude higher than the limit inferred from loss in the array junctions (see~\ref{app:half-flux-compare}) and values observed in prior work~\cite{serniak2018hot, Pop2014Coherent}. 
Thus, absent an independent measurement of $x_\text{qp}$, we assume a negligible value of $x_\text{qp} = 1\times10^{-7}$ in Fig.~\ref{fig:t1-flux-baseline}a.

We note that other sources of phase-coupled noise, such as inductive loss described by Johnson-Nyquist current noise and QP tunneling across the array junctions, are difficult to distinguish from $1/f$-like flux noise with frequency dependence alone (see~\ref{app:models}). However, inductive loss lacks a candidate microscopic model, and temperature dependence measurements (discussed in Section~\ref{sec:T-dependence}) allow us to rule out QP tunneling in the array junctions (see~\ref{app:half-flux-compare}). Therefore, given that our spin-locking and spin-echo measurements are consistent with $1/f$-like flux noise, we proceed to model $T_1$ near half-flux with flux noise.
Additionally, we verify that radiation to the flux and charge lines (not shown) is negligible (see~\ref{app:models}).

So far, we have only considered transitions within the qubit (two-level) subspace. However, it is apparent from the gap between the models and the measured $T_1$---particularly at intermediate flux biases---that the model remains incomplete (light dash-dotted lines in Fig.~\ref{fig:t1-flux-baseline}a). 
To address the discrepancy, we consider the effect of the higher fluxonium levels on decoherence.
It is particularly pertinent to consider this effect for qubit designs with low-energy transitions close to the thermal energy scale, $hf_{ij} \approx k_BT$, and a lack of parity-protected transitions.
We use a rate matrix model to numerically evaluate the population transfer between the first $N$ fluxonium levels, whose populations are encoded in a length-$N$ vector $\vec{p}(t)$ (see~\ref{app:heating} for details).
The time dynamics are described by the equation $\partial_t\vec{p}(t) = B \vec{p}(t)$, where
% % 
\begin{equation}
    B = \begin{pmatrix} 
    -\sum\limits_i\Gamma_{0\rightarrow i} & \Gamma_{1\rightarrow 0}  & \cdots & \Gamma_{N\rightarrow 0}  \\ 
    \Gamma_{0\rightarrow 1}  &  - \sum\limits_i\Gamma_{1\rightarrow i} & \cdots & \vdots\\
    \vdots & \vdots & \ddots & \Gamma_{N\rightarrow N-1}    \\
    
    \Gamma_{0\rightarrow N}  & \Gamma_{1\rightarrow N}  & \cdots &  - \sum\limits_i \Gamma_{N\rightarrow i} 
    \end{pmatrix} \label{eq:B-matrix-main}
\end{equation}
is an $N\times N$ matrix, and the transition rates from $\ket{i}$ to $\ket{j}$, $\Gamma_{i\rightarrow j}$, are calculated by Fermi's golden rule.
% % 
%%%%%%%%%%%%%%%%%%%%%%%%%%%%%%%%%%%%%%%%%%%%%%%%%%%%%%%%%%
Due to the matrix-element structure of the fluxonium, the rates of heating-induced transitions to short-lived, non-computational states are non-negligible. 
We use $N=6$ levels---at which point the result converges---and plot the decay rates for the same models considered above (dark dashed lines in Fig.~\ref{fig:t1-flux-baseline}a).
We find that at all flux biases, the time evolution is nearly exponential, and the net result is a corrected depolarization rate $\Gamma_1^\text{eff}$, which represents the decay rate of an eigenmode of matrix $B$ (see~\ref{app:heating} for more details).
Uncertainties in the predicted $T_1$ due to the potential of populating non-computational levels (contributing, in principle, to deviations from exponential behavior and readout errors) are within percent level at all flux biases, with the largest uncertainties of $1-7\%$ in regions near avoided crossings in the spectrum (see~\ref{app:heating}).
When including these higher-level transitions, the total $T_1$ prediction of our model (dark solid line in Fig.~\ref{fig:t1-flux-baseline}a) aligns significantly better with the data, with the strongest contribution to the correction coming from dielectric loss (purple). 
(We note that resonant-like dips appear in the combined population evolution due to Purcell-induced heating near resonator-fluxonium crossings; e.g., at $|\delta\Phi_\text{ext}| \approx 0.06$, the resonator intersects the $\ket{0}\leftrightarrow\ket{3}$ transition.)

While the contribution of different noise sources and higher levels to the $T_1$ preclude us from extracting a single noise power spectrum from the data, normalizing the measured depolarization rates by the square of a transition matrix element provides further insight into the frequency structure of the noise. 
We plot the corresponding normalized quantity, $\Gamma_1 / |\bra{0} \hat{D}_\lambda \ket{1} |^2$, for $\hat{D}_\lambda = \hat{\phi}$ and $\hat{n}$ in Figs.~\ref{fig:t1-flux-baseline}b and ~\ref{fig:t1-flux-baseline}c, respectively.
One can derive a relation between the matrix elements of the conjugate operators $\hat{\phi}$ and $\hat{n}$, $|\langle i | \hat{n} | j\rangle| = \frac{\hbar\omega_{ij}}{8 E_C} |\langle i | \hat{\phi} | j\rangle|$, which results in the two plots differing by a factor proportional to $f_{01}^2$.
% Due to the conjugate matrix element relation, $|\langle i | \hat{n} | j\rangle| = \frac{\hbar\omega_{ij}}{8 E_C} |\langle i | \hat{\phi} | j\rangle|$, the two quantities are related by a factor $\propto f_{01}^2$.
At the lowest frequencies, the flux noise $S_\phi \propto f_{01}^{-1}$ is consistent with a $1/f$-like dependence, while at higher frequencies, charge noise appears largely frequency-independent, $S_n \propto f_{01}^0$.
Although the behavior of $S_\phi \gtrsim 100~\mathrm{MHz}$ could be interpreted as super-Ohmic flux noise ($S_\phi \propto f_{01}^\gamma$, with $\gamma = 2$), the crossover to quantum (Nyquist) noise is only expected to occur at significantly higher frequencies,
% ~\footnote{Provided that the high-frequency cut-off of the fluctuations lies above the turnover due to temperature.}, 
$k_B T_\text{eff} / h \gtrsim 650~\mathrm{MHz}$~\cite{quintanaObservationClassicalQuantumCrossover2017, yanFluxQubitRevisited2016}. 
We note the drop in $S_n$ at $f_{01} \gtrsim 6~\mathrm{GHz}$, where the $\ket{0}\leftrightarrow\ket{1}$ transition changes in character from interwell to intrawell relaxation. 
The deviation of these points from the trend at lower frequencies
reflects the possible need to account for higher levels (as discussed above), or suggests that a single loss tangent may not describe the data across all flux biases.

% ---------------------------------------------------------------------------------------------------------
% Figure 3: Temperature Dependence
% ---------------------------------------------------------------------------------------------------------
\begin{figure*}[ht!]
    \centering
    \includegraphics[width=0.9\textwidth]{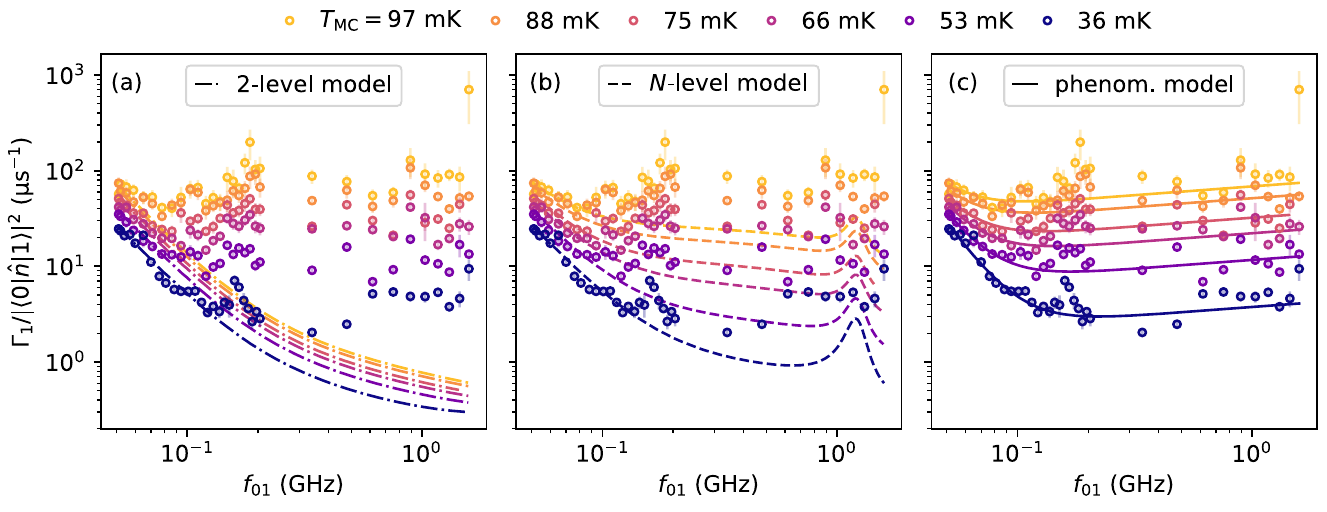}%
    \caption{\textbf{Temperature dependence and model comparison for fluxonium energy decay.} Temperature dependence of relaxation rates versus frequency, normalized by the magnitude squared of the charge matrix element, $\Gamma_1/\matel{n}^2$. 
    Each panel displays a candidate loss model, with the same data repeated across all three panels for clarity. A TLS identified at $f_{01} \approx 150~\mathrm{MHz}$ is disregarded in the modeling.
    (a) Two-level model: combination of loss mechanisms modeled with two fluxonium levels---$1/f^\alpha$ flux noise, dielectric loss, Purcell decay, and QP tunneling in the small junction (see main text). For $1/f^\alpha$ flux noise, a phenomenological temperature dependence is introduced: $A_\Phi(T) = A_\Phi(T_0) \cdot (T/T_0)$, with $T_0 = 36~\mathrm{mK}$. (b) $N$-level model: combination of same loss mechanisms as in panel (a) acting on $N=6$ fluxonium levels, where effective decay rates are calculated using the rate matrix in Eq.~\ref{eq:B-matrix-main}. 
    (c) Phenomenological power-law model: least-squares fit of the data to Eq.~\ref{eq:phenom-T-model}, which assumes power-law scaling with frequency and temperature (see main text).
    Measurements were taken asymmetrically about half-flux; at biases corresponding to equal frequencies, the points are averaged for clarity. 
    }
    \label{fig:T-sweep}
\end{figure*}

% ---------------------------------------------------------------------------------------------------------
% Temperature Dependence
% ---------------------------------------------------------------------------------------------------------
\section{Temperature Dependence}\label{sec:T-dependence}
Having identified $1/f^\alpha$ flux noise and dielectric loss as the predominant sources of energy decay, we proceed to investigate its temperature dependence at low qubit frequencies.
We measure $T_1$ as a function of flux bias while varying the temperature of the mixing-chamber stage up to $T_\text{MC} \approx 100~\mathrm{mK}$. 
In Figure~\ref{fig:T-sweep}, we plot the temperature-dependent energy decay rate as a function of frequency, which we normalize by the square of the charge matrix element,  $\Gamma_1/\matel{n}^2$. 
We attribute a peak in dissipation at approximately $150~\mathrm{MHz}$ to a stray TLS, which we exclude from the following analysis as we are primarily interested in the ensemble properties.

The measurements are compared against three candidate models, with the same data shown in each panel for clarity.
The first two models (Figs.~\ref{fig:T-sweep}a and~\ref{fig:T-sweep}b) comprise the same combination of loss mechanisms detailed in Section~\ref{sec:T1-baseline}: $1/f^\alpha$ flux noise, dielectric loss, QP tunneling in the small junction, and Purcell decay.
The first (second) panel shows the two-level ($N$-level) approach, where the latter is evaluated with the rate matrix in Eq.~\ref{eq:B-matrix-main}.
To model the temperature dependence of the flux noise, we include a phenomenological $A_\Phi(T) = A_\Phi(T_0) \cdot (T/T_0)$ with $T_0 = 36~\mathrm{mK}$.
In applying these models, we fit the base-temperature data to the two-level model to extract loss parameters at $T_\text{eff} = 36~\mathrm{mK}$ (obtaining $\epsilon = 0.45$ to describe the frequency dependence of the dielectric loss for this dataset), then evaluate them at each temperature without additional free parameters. 
% ~\footnote{The DR underwent a thermal cycle between this and the preceding measurement, so a potential rearrangement of defects could explain the difference in $\epsilon$ between cooldowns.}
For simplicity, we use $T_\text{eff} = T_\text{MC}$ and
$T_\text{res} = \text{max}(T_\text{MC},~70~\mathrm{mK})$ for Purcell, reflecting a low-temperature saturation of the resonator temperature~\footnote{The observation that $T_\text{eff} = T_\text{MC}$ describes our low-frequency data well, in contrast to the estimated low-temperature saturation of $T_\text{eff} = 50~\mathrm{mK}$ away from half-flux, may suggest that the low- and high-frequency defect baths are described by different effective temperatures}.
As an additional, third, comparison, we consider a purely phenomenological power-law model in Fig.~\ref{fig:T-sweep}c, which we detail later.

We start by discussing the temperature dependence of the low-frequency flux noise, which is largely unaffected by the higher levels due to the large qubit anharmonicity near half-flux (Figs.~\ref{fig:T-sweep}a and~\ref{fig:T-sweep}b). 
By fitting a temperature-dependent $A_\Phi$ to the $T_1$ data, we find an approximately linear scaling $A_\Phi \propto T_\text{MC}$ (see~\ref{app:T-dependent-dephasing}).
We confirm in a separate temperature sweep that the 
echo pure-dephasing rates are consistent with $A_\Phi$ following a similar trend, indicating that the same noise source underlying the qubit dephasing is likely responsible for  energy decay (see~\ref{app:T-dependent-dephasing}). 
Notably, the linear dependence of flux noise on temperature is also consistent with a model of inductive loss described by Johnson-Nyquist noise (see~\ref{app:models})~\cite{Hazard2019Nanowire, Zhang2021Universal}, which predicts temperature scaling proportional to $\coth(\hbar\omega/2k_BT) \sim k_BT/\hbar\omega$ when $\hbar\omega \ll k_BT$. 
% Although a candidate microscopic model behind the inductive loss mechanism is unclear, it has been speculated that flux and inductive noise may derive from a common source~\cite{sendelbachComplexInductanceExcess2009}.
In contrast, prior studies of the flux noise in SQUIDs and flux qubits have observed little-to-no dependence in the flux noise with temperature below approximately 100 mK~\cite{wellstoodLowFrequencyNoise1987, sendelbachMagnetismSQUIDsMillikelvin2008, antonMagneticFluxNoise2013, yanSpectroscopyLowfrequencyNoise2012, quintanaObservationClassicalQuantumCrossover2017}.
We remark that characterization of the quasi-static dephasing noise with repeated Ramsey measurements reveals anomalous Lorentzian features atop the $1/f$-like trend (see~\ref{app:low-freq-flux-noise}), similar to those observed in Ref.~\cite{rower2023evolution} with applied magnetic fields.

Now we turn to the higher frequencies, where it is evident from Fig.~\ref{fig:T-sweep}a that the two-level models fail to capture the temperature dependence.
The $N$-level model (Fig.~\ref{fig:T-sweep}b) shows significantly better agreement with the data, suggesting that higher-level transitions play an important role in the observed behavior.
The remaining discrepancy (seen in the offsets of the $T$-dependent curves from the data in Fig.~\ref{fig:T-sweep}b) may be related to a temperature-dependent loss tangent, but could also suggest an elevated bath temperature relative to the MC stage.

Given the challenge of explaining our high-frequency data by modeling individual loss channels, we turn to a purely phenomenological power-law model for both the low- and high-frequency noise spectra (\fref{fig:T-sweep}c). In terms of the normalized decay rate, we express this as
%
% \begin{align}
% \begin{split}
%     \frac{\Gamma_1^\text{phen.}}{|\langle 0 |\hat{n} |1\rangle|^2} 
%     &= \frac{A}{(\omega/\omega_0)^{{(2+\alpha)}}} 
%     \left( \frac{T}{T_{{0}}} \right)^{{\beta_1}}
%     % &\quad 
%     + B  \left(\frac{\omega}{\omega_1}\right)^\gamma \left( \frac{T}{T_{{0}}} \right)^{{\beta_2}}, \label{eq:phenom-T-model}
% \end{split}
% \end{align}
\begin{equation}
    \frac{\Gamma_1^\text{phen.}}{|n_{01}|^2}  = \left( \frac{8E_CE_L}{\hbar^2 \phi_0} \right)^2 A\omega_{01}^{-\alpha-2}T^{\beta_1} + \left( \frac{4E_C}{e\hbar} \right)^2 B\omega_{01}^\gamma T^{\beta_2}, \label{eq:phenom-T-model}
\end{equation}
where we have effectively defined flux and charge noise power spectral densities with a power-law dependence on frequency and temperature: $S_\Phi(\omega) = A\omega^{-\alpha}T^{\beta_1}$ and $S_Q(\omega) = B\omega^\gamma T^{\beta_2}$, and we introduce the shorthand $|n_{01}| \equiv |\langle 0 |\hat{n} |1\rangle|$.
Note that $\alpha = 1$ corresponds to $1/f$ flux noise.
Fitting Eq.~\ref{eq:phenom-T-model} simultaneously to the full dataset with $T = T_\text{MC}$ (excluding the region of the TLS), we extract $\alpha = 1.5$, $\beta_1 = 0.32$, $\gamma = 0.19$, and $\beta_2 = 2.9$ (see \fref{fig:T-sweep}c).
The low-frequency parameters are impacted by the fact that the second term in Eq.~\ref{eq:phenom-T-model} goes to zero as $\omega \rightarrow 0$, whereas a more realistic model comprising flux noise and dielectric loss has contributions from both mechanisms at low frequency. At high frequency, on the other hand, the contribution from flux noise falls off, and the remaining loss can be largely attributed to charge noise.
This component displays a temperature dependence of approximately $T^{3}$, accompanied by a sub-Ohmic frequency scaling of $\omega^{0.2}$. 

%%%%%%%%%%%%%%%%%%%%%%%%%%%%%%%%%%%%%%%%%%%%%%%%%%%%%%%%%%%%%%%%
% 
To address the $T^{3}$ scaling of the high-frequency charge noise, we consider two ways such a power-law dependence could arise within the standard tunneling model (STM) for TLS~\cite{Phillips1987TwoLevelStates} (see~\ref{app:TLS} for more details).
We note that the following discussion also pertains to $T^{\beta_2}$ scaling with $\beta_2 \neq 3$, as the potential effect of higher-energy levels complicates the extraction of this exponent. 
The STM offers a description of non-interacting tunneling defects in amorphous oxides that couple to electromagnetic and strain fields~\cite{Phillips1987TwoLevelStates}.
The confinement of such modes in low-dimensional structures also shapes the dissipation from these defects~\cite{Behunin2016Dimensional},
which generally arises from two mechanisms: (1) resonant and (2) relaxation absorption~\cite{Phillips1987TwoLevelStates, Behunin2016Dimensional}. 
Resonant absorption, which involves the absorption of a photon by a defect with matching energy, has been well studied in the context of superconducting microwave resonators and qubits at millikelvin temperatures~\cite{martinisDecoherenceJosephsonQubits2005, gaoPhysicsSuperconductingMicrowave, McDermott2009Materials, Burnett2016Analysis, Wang2020Materials, crowley2023disentangling}. 
The loss tangent resulting from resonant absorption in a TLS ensemble scales as $\tan\delta_\text{TLS, res} \propto \tanh(\hbar\omega/2k_BT)/\sqrt{1+\bar{n}/n_c}$, where $\bar{n}$ is the average circulating photon number; $n_c \propto (\tau_1\tau_2)^{-1}$ is a critical photon number that characterizes the saturation of the process, and $\tau_1$ and $\tau_2$ are the defect energy relaxation and dephasing times, respectively~\cite{Phillips1987TwoLevelStates}.
For large photon numbers, a defect lifetime scaling with temperature as $\tau_1^{-1} \propto T^{\beta_2}$ with $\tau_2 = 2\tau_1$ could plausibly result in a power-law dependence in the qubit depolarization rate $\Gamma_1^\text{TLS, res} \propto T^{\beta_2} |\langle 0|\hat{n}|1 \rangle |^2$.
It is unclear what would give rise to such a power-law scaling; however, we note that while a phonon-limited defect lifetime has the dependence $\tau_1^{-1} \propto \coth(\hbar\omega/2k_BT)$ (as is typically assumed for TLS in glasses~\cite{phillipsTunnelingStatesAmorphous1972, Behunin2016Dimensional}), TLS lifetimes have also been experimentally observed to follow a parabolic temperature dependence~\cite{lisenfeldMeasuringTemperatureDependence2010}, suggesting that other temperature scaling behaviors are plausible. 

Relaxation absorption, on the other hand, involves the longitudinal modulation of defect energies by off-resonant photons, resulting in instantaneous shifts of the defect thermal equilibrium population and subsequent relaxation~\cite{Phillips1987TwoLevelStates, Behunin2016Dimensional, Wollack2021LossLN}.
% to the new equilibria. 
To equilibrate, the defect is typically assumed to emit the energy into a phonon mode~\cite{Phillips1987TwoLevelStates}. 
In contrast to resonant absorption, this process is not saturable and is exponentially suppressed at high frequencies relative to the thermal energy scale, $\hbar\omega \gg k_BT$. 
At low frequencies, it can lead to a temperature scaling of the dissipation $\tan\delta_\text{TLS, rel} \propto T^d$, where $d$ characterizes the dimension of the phonon bath. 
This temperature scaling has been observed in nanomechanical resonators ($d \approx 2.3$-$2.4$~\cite{MacCabe2020Nano, Wollack2021LossLN}), in the critical current noise of SQUIDs ($d=2$)~\cite{wellstood2004flicker}, and in the high-temperature dielectric loss of silicon nitride~\cite{Mittal2024Annealing} ($d = 2$). 
Linking this to our observation, a $d=3$ scaling could suggest consistency with a bulk geometry. 
However, the corresponding qubit relaxation rate scales as $\Gamma_1^\text{TLS, rel} \propto  \omega^{-1} T^d \coth(\hbar\omega/2k_BT) |\langle 0|\hat{n}|1 \rangle |^2$, which does not agree well with the frequency dependence of our data. 
We suggest further studies to understand the possible relevance of relaxation-type TLS damping in low-frequency qubits, such as investigating the crossover regime from low to high frequency, the temperature dependence of the mode frequency, and the time dependence of the relaxation times.
% 

% ---------------------------------------------------------------------------------------------------------
% Figure 4: Magnetic field dependence:
% ---------------------------------------------------------------------------------------------------------
\begin{figure*}
    \centering
    \includegraphics[width=\textwidth]{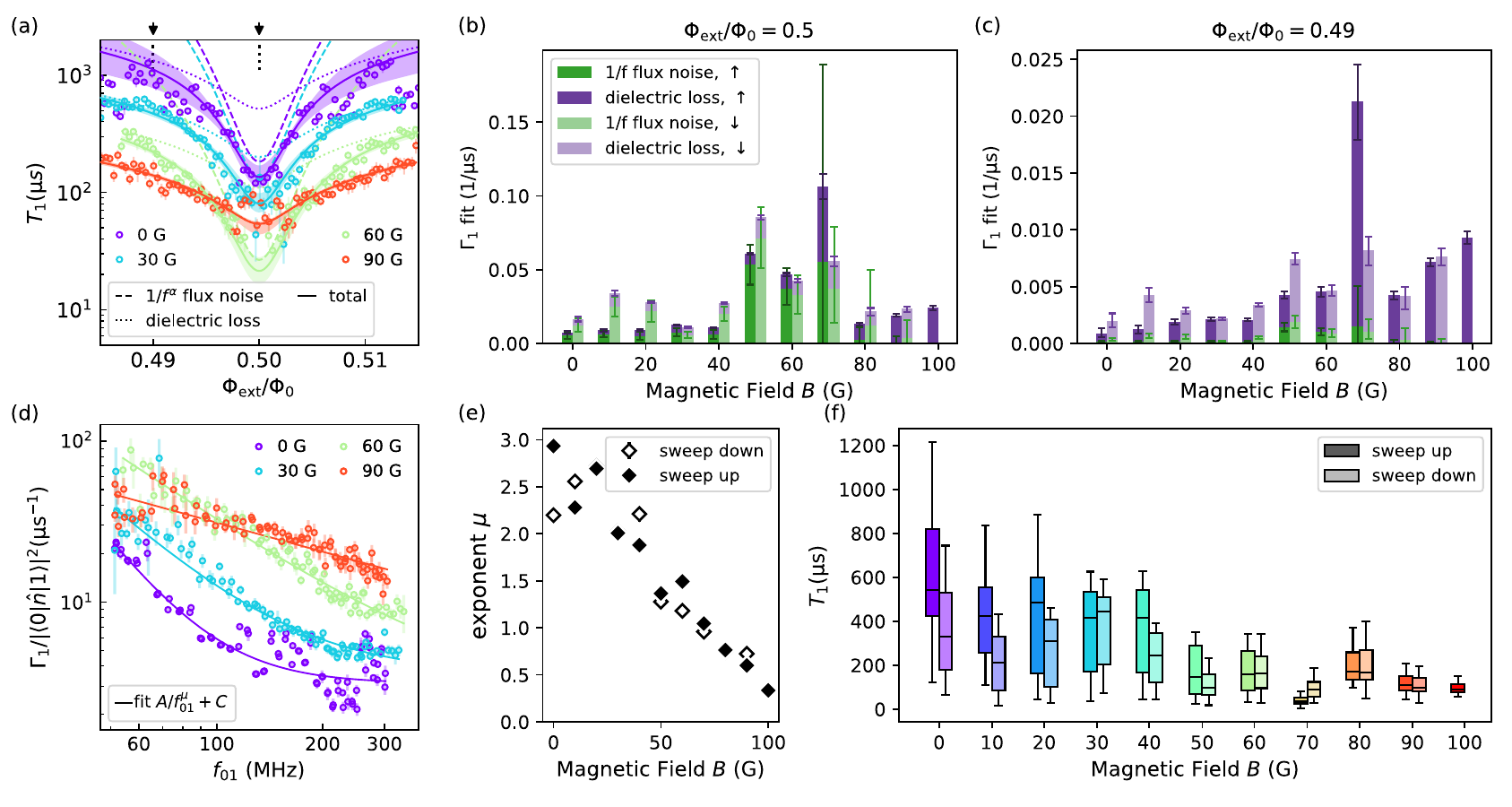}
    \caption{\textbf{In-plane magnetic-field dependence of fluxonium $T_1$.} (a) Measured $T_1$ versus flux bias at four representative in-plane magnetic fields. Full dataset is shown in~\ref{app:more-B-field-data}; these traces correspond to increasing values of magnetic field (sweep up). 
    Data are fit to a combination of $1/f^\alpha$ flux noise and dielectric loss with fixed parameters $\alpha = 0.62$ and $\epsilon = 0.31$ obtained from fits at $B=0~\mathrm{G}$. Shaded regions represent 95\% confidence intervals obtained from empirical bootstrapping with $n=1000$ samples (see main text). At fields $B < 20~\mathrm{G}$, we constrain the fit to $|\delta\Phi_\text{ext}| \leq 5~\mathrm{m\Phi_0}$ to avoid the TLS structure away from half flux. (b) Fitted values of $\Gamma_1$ at $\Phi_\text{ext}/\Phi_0 = 0.5$ and (c) $\Phi_\text{ext}/\Phi_0 = 0.49$. Estimated contributions from flux noise (green) and dielectric loss (purple) are indicated by the height of the stacks. 
    Darker (lighter) colors indicate the upwards (downwards) field sweep, and error bars represent 95\% confidence intervals of the bootstrap distribution. (d) Same data as in panel (a) plotted as $\Gamma_1/\matel{n}^2$, overlaid with fits to $y=A/f_{01}^\mu + C$. (e) Extracted fit exponent $\mu$ as a function of magnetic field (all parameters can be found in~\ref{app:more-B-field-data}). (f) Box-and-whisker plot of $T_1$ distribution as a function of magnetic field. Lines indicate the median, boxes represent the inner two quartiles, and whiskers denote the minimum/maximum data points taken in the flux-bias region shown in panel (a).}
    \label{fig:T1-B-sweep}
\end{figure*}

% ---------------------------------------------------------------------------------------------------------
% Magnetic field dependence
% ---------------------------------------------------------------------------------------------------------
\section{Magnetic-field Dependence}\label{sec:B-dependence}
As our final experiment, we investigate the in-plane magnetic-field dependence of $T_1$ with applied fields of up to $B = 100~\mathrm{G}$ (Fig.~\ref{fig:T1-B-sweep}).
We focus on the frequency range of $52~\mathrm{MHz} \leq f_{01} \leq 300~\mathrm{MHz}$, shown in Fig.~\ref{fig:setup}d,
where the qubit depolarization under baseline conditions is relatively well described by $1/f^\alpha$ flux noise and dielectric loss (Fig.~\ref{fig:t1-flux-baseline}a, inset).
We sweep the applied field in steps of $10~\mathrm{G}$ from $B_\text{min} = 0~\mathrm{G}$ to $B_\text{max} = 100~\mathrm{G}$ and back down, maintaining the direction of the sweep between $B_\text{min}$ and $B_\text{max}$ to avoid uncontrolled hysteretic effects. 
At each field, we locate the voltage bias corresponding to the degeneracy point, recalibrate readout parameters, and measure $T_1$ as a function of flux bias. 
Data are taken at $T_\text{MC} \approx 45~\mathrm{mK}$. 
We observe a decreasing $E_J$ with increasing field following a Fraunhofer pattern~\cite{schneiderTransmonQubitMagnetic2019}, and find that the change in Hamiltonian parameters with field results in only percent-level changes to $f_{01}$, $\matel{\phi}$, and $\matel{n}$ (see~\ref{app:B-field-qb-params}).

We start by fitting the $T_1$ data to a combination of $1/f^\alpha$ flux noise and dielectric loss at each magnetic field. Figure~\ref{fig:T1-B-sweep}a shows representative traces at four magnetic fields, while the complete dataset is provided in~\ref{app:more-B-field-data}. We exclude Purcell and QP tunneling from our analysis, as their contributions near the degeneracy point are negligible and expected to remain so across the magnetic field sweep. 
We further elect to use the two-level models for simplicity, while confirming that the multi-level model does not affect our qualitative conclusions. 
Informed by the measurements in the absence of applied field, we use the same parameters as those in Fig.~\ref{fig:t1-flux-baseline}a to constrain our fits: $\tan\delta_C^0 = 4\times10^{-6}$ and $\alpha = 0.62$ yields values $A_\Phi = (0.23~\mathrm{\mu\Phi_0})^2$ and $\epsilon = 0.31$ at $B = 0~\mathrm{G}$, consistent with those in Fig.~\ref{fig:t1-flux-baseline}a. To avoid over-fitting, we hold the values of $\alpha$ and $\epsilon$ fixed and fit $A_\Phi$ and $\tan\delta_C^0$ at each magnetic field.
While in principle the noise exponents may (and likely do) change with field (see, e.g.,~\cite{rower2023evolution}), our qualitative conclusions below remain unchanged for fixed values between $0.5 < \alpha < 1.5$ and $0 < \epsilon < 0.7$ (see~\ref{app:more-B-field-data} for further discussion).
We note that more dramatic changes in these exponents with field could affect our conclusions, motivating follow-up studies to independently characterize the charge and flux noise spectra~\cite{Rower2024Purity}.

To obtain error bars on the fits, we use empirical bootstrapping~\cite{andrae2010error}, where points are sampled from the dataset with replacement to generate multiple ``bootstrapped datasets'' from a single measured dataset. By generating $n = 1000$ such bootstraps, we perform a least-squares fit to each set to build up a distribution of parameters, from which we obtain $95\%$ confidence intervals of the fits, indicated by the shaded regions in Fig.~\ref{fig:T1-B-sweep}a. 

To understand the breakdown of flux noise and dielectric loss as a function of field, we compute the contributions to $\Gamma_1$ from each model. 
We look at two representative bias points:
$\Phi_\text{ext}/\Phi_0 = 0.5$---at half-flux, where $\matel{\phi}$ is large---and $\Phi_\text{ext}/\Phi_0 = 0.49$---off half-flux, where $\matel{\phi}$ is small (Figs.~\ref{fig:T1-B-sweep}b and~\ref{fig:T1-B-sweep}c, respectively). 
At the degeneracy point ($f_{01} = 52~\mathrm{MHz}$), $1/f^\alpha$ flux noise dominates the decay rate for low in-plane fields, as expected from the correspondingly large $\Gamma_1$ susceptibility to phase-coupled noise.  At intermediate fields, $50 \lesssim B \lesssim 80~\mathrm{G}$, a broad peak in the flux noise appears, followed by a significant decrease at $B \gtrsim 80~\mathrm{G}$, where the flattening of the $T_1$ variation with flux-bias suggests a transition in the dominant loss mechanism from flux noise to dielectric loss. 
We note that this jump resembles a similar feature in Ref.~\cite{rower2023evolution}: although~\cite{rower2023evolution} probed a significantly higher frequency band, a similar jump in $\Gamma_1$ appeared between 40 and 80 G.

On the other hand, off the degeneracy point ($f_{01} = 200~\mathrm{MHz}$), where the suppression of $\matel{\phi}$ leads to predominantly charge-coupled noise, $\Gamma_1$ largely tracks the variation of dielectric loss with field (Fig.\ref{fig:T1-B-sweep}c).
This trend appears to suggest that dielectric loss \textit{increases} with applied magnetic field. 

We now examine the data through an alternative lens that emphasizes the frequency structure of the noise. Once again, we normalize the decay rate to the square of the charge matrix element (Fig.~\ref{fig:T1-B-sweep}d). 
Motivated by the monotonically decreasing trend with frequency, we fit the measured data at each field to the sum of an inverse-frequency and white noise component: $y = A / f_{01}^{\mu} + C$. 
The fit exponent $\mu (B)$ (Fig.~\ref{fig:T1-B-sweep}e) decreases with field from $\mu \approx 3$ at 0 G (consistent with $1/f$-like flux noise) to $\mu \approx 0.3$ at 100 G (consistent with $1/f$-like charge noise with spectrum slope $\mu \approx 0.3$~\cite{Nakamura2002}).
This trend also suggests a transition in the dominant loss mechanism from flux noise at low fields to charge noise at high fields, consistent with our conclusion from the previous fits to individual loss models. 
This qualitative trend was reproduced in another field sweep (see~\ref{app:more-B-field-data}). 

Finally, we plot a summary of the measured $T_1$ values to highlight their non-monotonic evolution with in-plane magnetic fields. 
A box-and-whisker plot (\fref{fig:T1-B-sweep}f) presents the evolution of the median, inner two quartiles, and spread of the $T_1$ data taken within $|\delta\Phi_\text{ext}| \leq 15~\mathrm{m\Phi_0}$.
Both the medians and distributions decrease non-monotonically with field, while the spread of $T_1$ values becomes more narrow.
This narrowing may be explained by the smoothing of resonant TLS features from increased qubit dephasing (broadening the spectrum linewidth), as well as the transition to loss dominated by a mechanism with weaker flux dependence (here, dielectric loss). 
We briefly remark on the slight hysteresis in the measured $T_1$ values at low fields, which are typically suppressed when the field is swept back down, compared to the initial upward sweep.

We now discuss several implications of this result, first considering alternative sources of field-induced noise. 
Studies of field-cooled devices suggest that the loss associated with vortex generation should monotonically decrease with field, in contrast to our observed trends~\cite{Song2009Reducing, Bothner2012Magnetic} (although we do not field-cool our device).
We also note that bounding the superconducting-gap suppression with field allows us to exclude the thermal generation of QPs (\ref{app:B-field-qb-params}), 
although the onset of another QP generation mechanism with field (e.g., related to the gap disorder~\cite{graafTwolevelSystemsSuperconducting2020, meyer2020dynamical}) is worth investigating in future work. 

Notably absent from our data is a signature of resonant Zeeman-split $g = 2$ spins, which have been observed to limit the coherence of solid-state quantum devices~\cite{bluvsteinExtendingQuantumCoherence2019, paladinoNoiseImplicationsSolidstate2014, degraafDirectIdentificationDilute2017, Jayaraman2024, gunzler2025spin}. The lack of such a signal in our experiment may indicate the dominance of defect spin energy scales beyond or comparable to the applied Zeeman field, or low coherence of the spin bath. In contrast, electron spin-resonance (ESR) studies of native surface spins on high-field-resilient superconducting resonators~\cite{degraafDirectIdentificationDilute2017, borisovSuperconductingGranularAluminum2020} and qubits~\cite{gunzler2025spin} have observed resonant dips in $T_1$ at the Zeeman frequency for field strengths above 2000 G ($\gg B_\text{max}$), with linewidths on the order of 100 MHz.
Indeed, significant evidence points to the need to account for spin-spin interactions~\cite{burnettEvidenceInteractingTwolevel2014, antonMagneticFluxNoise2013}, for which strengths in the range of 200-800 MHz have been proposed~\cite{sendelbachMagnetismSQUIDsMillikelvin2008, lisenfeld2015observation, rower2023evolution}.
In this regime, inter-defect coupling dominates over the Zeeman energy scale, and the uncoupled-spin picture is not justified. 
The apparent transition to dielectric-loss-limited coherence at high fields also limits the sensitivity of our qubit to such features.
To combat this difficulty, a gradiometric fluxonium~\cite{gunzler2025spin, Sun2023FluxoniumLossMech, quintanaObservationClassicalQuantumCrossover2017} would serve to decouple the tuning of qubit frequency and matrix elements.

The apparent increase (within our analysis framework) in dielectric loss with field is more perplexing, as the defects comprising the dominant charge noise TLS bath are not typically associated with a magnetic moment. 
However, correlations between charge noise reduction and magnetic surface spin removal have pointed to a possible link between the two baths~\cite{degraafSuppressionLowfrequencyCharge2018, Jayaraman2024}. 
It was recently shown that weak magnetic fields could reduce the dielectric loss induced by implanted boron acceptors due to their spin-orbit structure~\cite{Zhang2024AcceptorInduced}, and while intrinsic Si is unlikely to host enough acceptors to dominate dielectric loss, the possible relevance of spin-orbit coupling---for example, associated with dangling-bond defects in surface oxides---has been considered~\cite{albertoPersonalComm}.
Our results, consistent with charge noise responding to applied magnetic fields, provide further evidence for a potential link between flux and charge noise, motivating the development of theories describing magnetic-field-dependent dielectric loss.

% ---------------------------------------------------------------------------------------------------------
% Conclusion
% ---------------------------------------------------------------------------------------------------------
\section{Conclusion}\label{sec:conclusion}
We characterize the temperature, magnetic-field, and flux-bias dependence of $T_1$ in a low-frequency fluxonium qubit, probing trends in the flux and charge noise that may help clarify the underlying noise mechanisms.
As the fluxonium energy-level transitions (including and beyond $\ket{0}\leftrightarrow\ket{1}$) can be close to the thermal energy, accounting for multiple levels gives a better description of the loss. 
At low frequencies, we find the flux-noise spectrum to be temperature dependent with $A_\Phi \propto T$, in contrast to prior studies that observed temperature-independent flux noise.
At higher frequencies, where we measured between $f_{01} \approx 100~\mathrm{MHz}$ and 1.6 GHz, the charge noise appears to increase substantially faster with temperature than predicted by a simple two-level dielectric loss model (not accounting for an intrinsic temperature dependence of the loss tangent).
When fit to a phenomenological power-law model, an approximately $T^{3}$ scaling fits the data best. 
Finally, the in-plane magnetic-field dependence suggests that the dielectric loss increases with applied field, pointing to a potential magnetic response of the electrically coupled defects---possibly indicating a microscopic link between the environmental charge and spin baths.
We anticipate these results may spur the development of microscopic theories of coherence-limiting defects that corroborate the observed temperature and magnetic field trends, leading to a deeper understanding of the intrinsic noise mechanisms that limit superconducting qubits.

% ---------------------------------------------------------------------------------------------------------
% Acknowledgments
% ---------------------------------------------------------------------------------------------------------
\section{Acknowledgments}\label{sec:acknowledgments}
We appreciate valuable conversations with José Alberto Nava Aquino, Jorge Marques, Sarang Mittal, Jeff Gertler, Mallika Randeria, and Kunal Tiwari.
We also gratefully acknowledge the open-source software package \href{https://scqubits.readthedocs.io/en/v4.1/}{scQubits}~\cite{Chitta2022Computer-aided, scQubits}, which was used for Hamiltonian diagonalization and matrix element calculation throughout this paper.
This material is based upon work supported by the U.S. Department of Energy, the Office of Science National Quantum Information Science Research Center’s Co-design Center for Quantum Advantage (Contract No. DE-SC0012704), and under Air Force Contract No. FA8702-15-D-0001.
L. A. and D. A. R. acknowledge support from the NSF Graduate Research Fellowship.
L. A. acknowledges support from the Laboratory for Physical Sciences Doc Bedard Fellowship. 
M.H. is supported by an appointment to the Intelligence Community Postdoctoral Research Fellowship Program at the Massachusetts Institute of Technology administered by Oak Ridge Institute for Science and Education (ORISE) through an interagency agreement between the U.S. Department of Energy and the Office of the Director of National Intelligence (ODNI).
The views and conclusions contained herein are those of the authors and should not be interpreted as necessarily representing the official policies or endorsements, either expressed or implied, of the U.S. Government.

% ---------------------------------------------------------------------------------------------------------
% Supplement / Appendix
% ---------------------------------------------------------------------------------------------------------
% \setcounter{figure}{0}
\setcounter{equation}{0}
\setcounter{section}{0}
\makeatletter 

\renewcommand{\thefigure}{\@arabic\c@figure}
\renewcommand{\thetable}{S\@arabic\c@table}
\renewcommand{\theequation}{S\arabic{equation}}
\renewcommand{\thesection}{Appendix \Alph{section}}
\renewcommand{\thesubsection}{\arabic{subsection}}

\makeatother
% \FloatBarrier

\section{Experimental setup} \label{app:setup}
\begin{figure}[hbt!]
    \centering
    \includegraphics[width=\columnwidth]{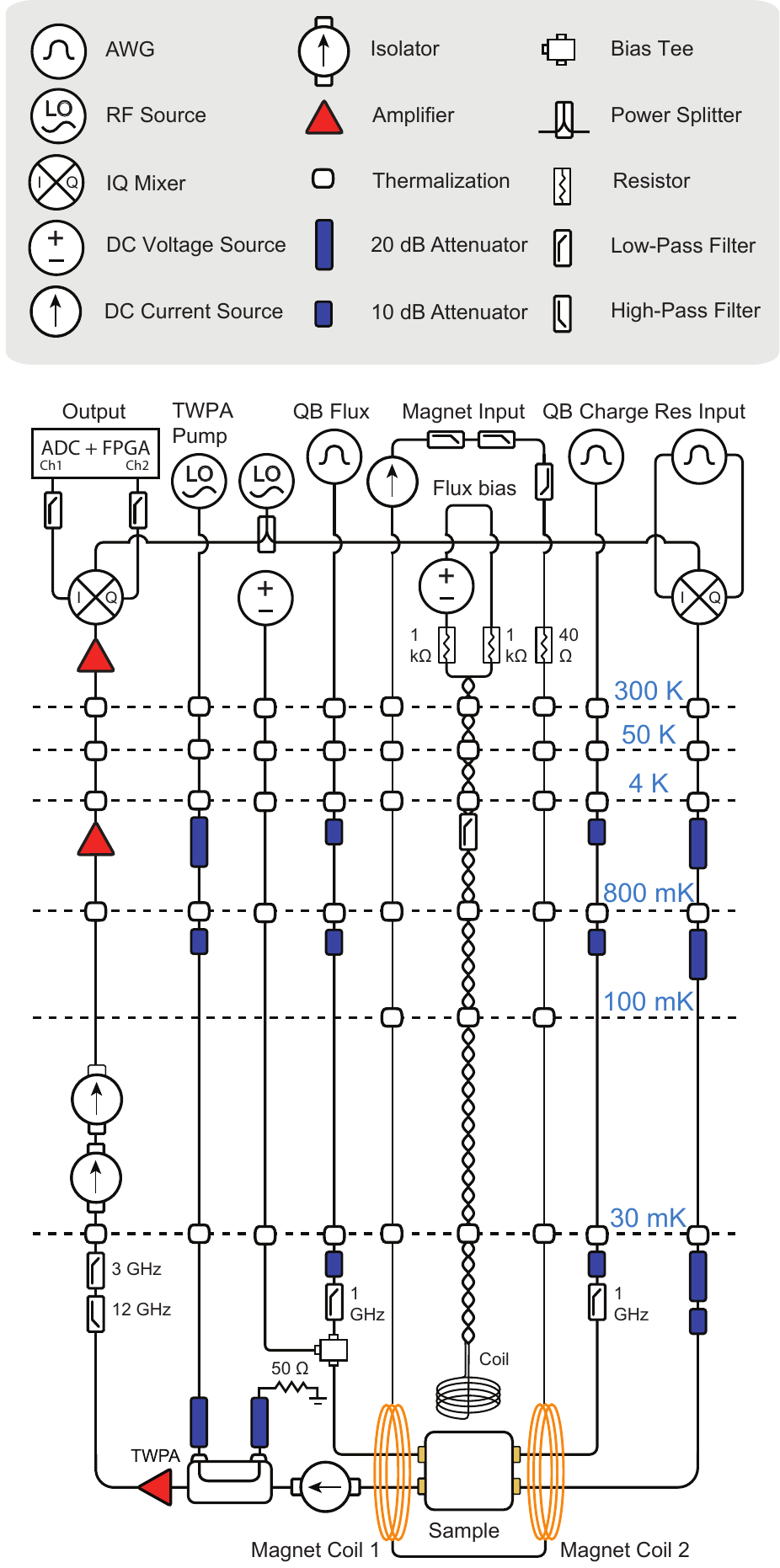}
    \caption{Wiring diagram of experimental setup.}
    \label{supfig:experimental-setup}
\end{figure}

We conducted our experiment in a Leiden CF-650 dilution refrigerator (DR) operating at a base temperature of approximately 30-40 mK. The experimental wiring is detailed in  Fig.~\ref{supfig:experimental-setup}. The sample was mounted within the magnet on a cold finger inside two layers of magnetic shielding comprising a superconducting aluminum shield surrounded by a Mu-metal encasing. 
A global out-of-plane flux bias was applied to the sample to tune the qubit frequency using a 500-turn bobbin mounted on the package lid. The flux bias was controlled using a differential voltage supplied by a QDevil QDAC with $1~\mathrm{k\Omega}$ room-temperature resistors on each channel.
Microwave lines with a total of 30 dB attenuation distributed across the stages were connected to the individual charge and flux bias lines on the chip; DC flux was combined with the radio-frequency (RF) flux signal at the mixing chamber stage using a bias tee. 
On-chip flux biasing was not used for the primary measurements of this experiment, so the flux line was terminated at room temperature (RT) except when measuring its mutual inductance to the qubit. 
Both the charge and flux lines had 1 GHz low-pass filters at the mixing-chamber stage when $T_1$ measurements were conducted. 
For low-frequency qubit control ($\lesssim$ 400 MHz, around the degeneracy point), AC control signals were directly synthesized with a Keysight M3202A arbitrary waveform generator (AWG) connected to the charge line.
A Rohde \& Schwarz SGS100 RF source supplied the high-frequency control drives ($>$ 400 MHz).
For readout, another Keysight M3202A AWG supplied a baseband signal, which was externally mixed with a local oscillator (Agilent E8257C) using IQ mixers for up- and down-conversion.
The readout output signals were amplified first by a Josephson traveling-wave parametric amplifier (JTWPA) pumped by a Holzworth HS9000 RF source, then by a high-electron mobility transistor (HEMT) amplifier at the 4 K stage, and finally a Stanford Research Systems SR445A amplifier at room temperature, before being digitized by a Keysight M3102A digitizer.
A Keysight M9019A chassis was used to synchronize and trigger the AWGs and digitizer, and a SRS FS725 rubidium clock provided global synchronization to all instruments. 

\begin{table}[h!]
\centering
\begin{tabular}{l l l} 
     \hline\hline
     Component & Manufacturer & Model \\ [0.5ex] 
     \hline
     Dilution fridge & Leiden & CF-650 \\ 
     RF source (readout) & Agilent & E8257C \\
     RF source (qubit) & Rohde \& Schwarz & SGS100 \\
     DC source (qubit) & QDevil & QDAC \\
     DC source (magnet) & Yokogawa & GS200 \\ 
     Control chassis & Keysight & M9019A \\ 
     AWG & Keysight & M3202A \\ 
     Digitizer & Keysight & M3102A \\
     [1ex] 
     \hline\hline
\end{tabular}
\caption{\textbf{Summary of control equipment.} The manufacturers and model numbers of the control equipment used for
the experiment.} \label{tab:control-equipment}
\end{table}

\subsection{Sample}

The sample comprised six uncoupled, floating fluxonium qubits with individual dispersively-coupled readout resonators coupled to a common transmission line. 
Only one qubit (our device-under-test) had both charge and flux lines wired to the external package connectors.
The sample was designed, fabricated, and packaged at MIT Lincoln Laboratory using their standard aluminum-on-silicon process, where the junctions (including the 151-junction array forming the superinductor, Fig.~\ref{fig:setup}b) were fabricated using electron beam lithography and double-angle shadow evaporation.
By design, the qubits had varying $E_J$ across the chip. 
The same set of designs has been characterized in~\cite{Azar2025}, and a similar design was used in~\cite{Randeria2024CQPS}. 
In the process used for these devices,
a cut-out was formed from each qubit to the chip edge to break up the ground plane and prevent screening currents surrounding the qubit when biasing. A TiPt patch (not shown)
% (shown in~\cite{Azar2025}) 
bonded the two sides of the cut-out for grounding.

\subsection{Magnet}

The magnet used in our experiment is similar to the one discussed in detail in Ref.~\cite{rower2023evolution}. 
It comprised two NbTi superconducting coils (hand-wound, 627 turns for each coil) with an inner radius of 15 mm, smaller than that of the coils used in~\cite{rower2023evolution}. 
The sample package was mounted between the two coils such that the rectangular junction loop had arms oriented parallel and perpendicular to the applied $B$-field.
Four Yokogawa GS200 DC sources in a parallel current-source configuration supplied an always-on bias current (with maximum current 705 mA at $B = 100~\mathrm{G}$).
The field per current at the sample location was 0.1418 G/mA.
Prior characterization of room-temperature filtering of the magnet wiring verified that noise from the power supply would not dominate qubit flux noise~\cite{rower2023evolution}.

\subsection{Temperature dependence}

For the temperature dependence measurements, $T_\text{MC}$ was set with an applied heating current at the mixing chamber stage. Datasets were collected after observing equilibration of repeated $T_1$ measurements at half-flux, typically within 45 minutes after a change in the current setting.
We note that the DR underwent a thermal cycle to room temperature between the baseline characterization (Fig.~\ref{fig:t1-flux-baseline}) and the temperature sweep (Fig.~\ref{fig:T-sweep}) so a potential rearrangement of charge defects with thermal cycling could explain the difference in the dielectric-loss frequency dependence $\epsilon$ observed between cooldowns.

% Res and qubit spec here
\begin{figure*}[th!]
    \centering
    \includegraphics[width=\textwidth]{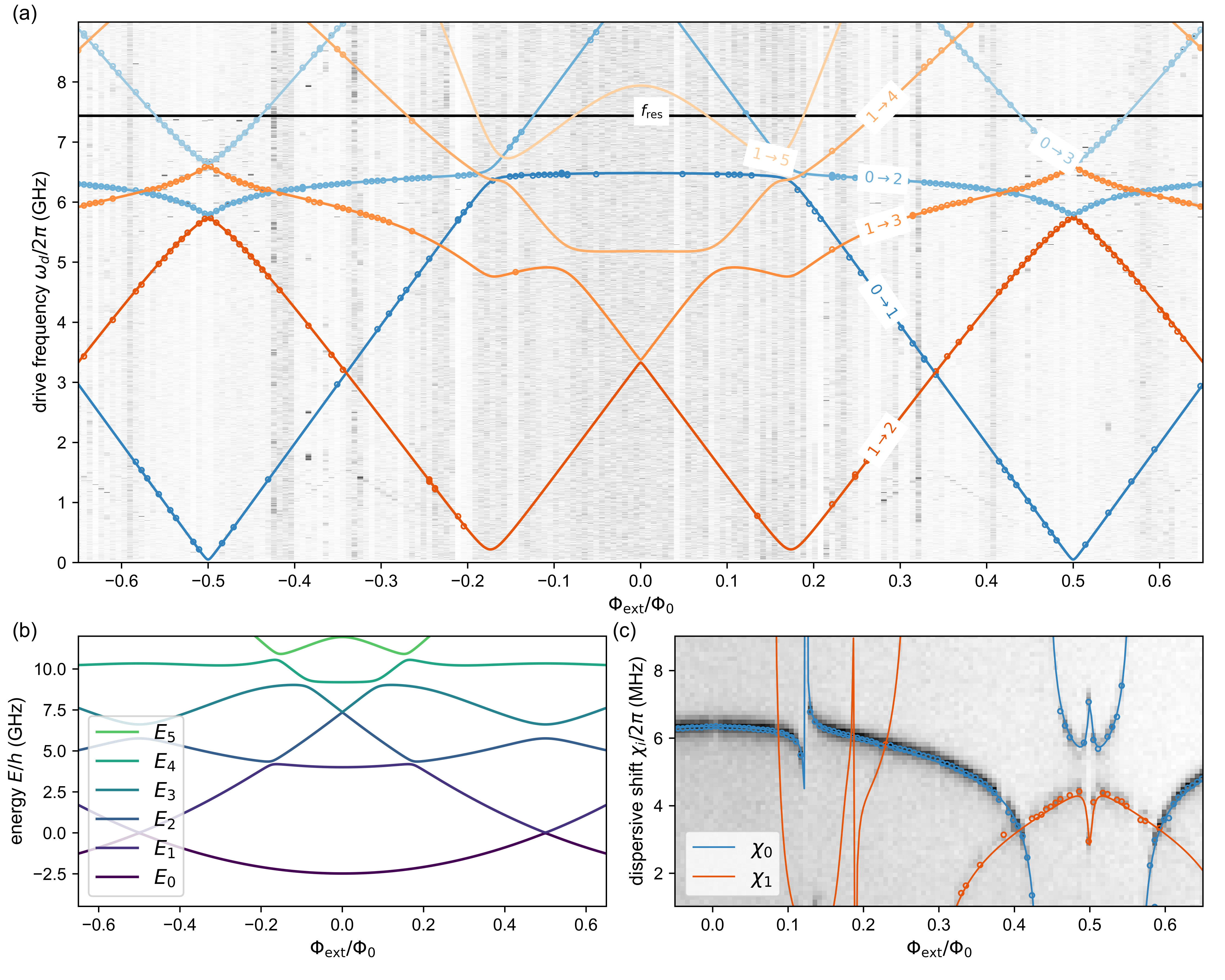}
    \caption{Qubit and resonator spectroscopy. (a) Qubit two-tone spectroscopy and fit to obtain Hamiltonian parameters. (b) Qubit energy spectrum without ground state energy subtracted off. (c) Resonator spectroscopy and fit to dispersive shifts $\chi_i$.}
    \label{supfig:qb-res-spec}
\end{figure*}

\section{Resonator and Qubit spectroscopy}\label{app:qb-res-spec}
% 
% \begin{figure*}[th!]
%     \centering
%     \includegraphics[width=\textwidth]{supp_figs/supp_qb_spec_wide_energy_levels_res_spec.png}
%     % {supp_figs/supp_qb_spec_wide_energy_levels_res_spec.pdf}
%     \caption{Qubit and resonator spectroscopy. (a) Qubit two-tone spectroscopy and fit to obtain Hamiltonian parameters. (b) Qubit energy spectrum without ground state energy subtracted off. (c) Resonator spectroscopy and fit to dispersive shifts $\chi_i$.}
%     \label{supfig:qb-res-spec}
% \end{figure*}

We obtained resonator and qubit spectra using standard single- and two-tone spectroscopy (\fref{supfig:qb-res-spec}), to which we applied a peak detector and extracted the system Hamiltonian parameters by numerical optimization.
The qubit-resonator Hamiltonian is given by
\begin{equation}
    \hat{H} = \hat{H}_\text{qb} + \hat{H}_\text{res} + \hat{H}_\text{int},
\end{equation}
where $\hat{H}_\text{qb}$ is given by Eq.~\ref{eq:hamiltonian}, $\hat{H}_\text{res} = \hbar\omega_\text{res} \hat{a}^\dagger \hat{a}$ and $\hat{H}_\text{int} = \hbar g\hat{n} (\hat{a} + \hat{a}^\dagger)$, where $\omega_\text{res} = 2\pi f_\text{res} = 2\pi \times 7.439~\mathrm{GHz}$ is the bare resonator frequency, $\hat{a}$ the resonator mode annihilation operator, and $g = 2\pi \times 124.6~\mathrm{MHz}$ the qubit-resonator coupling strength. To obtain $\omega_\text{res}$ and $g$, we fit the resonator dispersive shifts $\chi_i$ for qubit states $\ket{i} \in \{0,1\}$. 

\section{$T_1$ measurements, measurement-induced transitions, and estimating $T_\text{eff}$} \label{app:T1-measurements}
\begin{figure}[t!]
    \centering
    \includegraphics[width=\columnwidth]{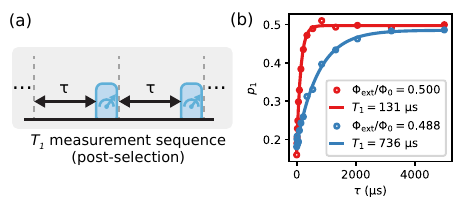}
    \caption{$T_1$ measurement sequence and example decay traces. (a) Pulse sequence for postselection. (b) Two example traces including exponential fit, heralding on the ground state $\ket{0}$.}
    \label{supfig:T1-sequence}
\end{figure}
Here we provide more detail on the $T_1$ measurement and discuss the implications of measurement-induced state transitions on the estimation of $T_\text{eff}$.
In our system, the MC temperature was higher than the qubit energy scale $f_{01} \lesssim k_B T_\text{MC} / h \approx 650~\mathrm{MHz}$ in the region of interest, resulting in significant equilibrium population of the first excited state.
State preparation was performed with two methods:
(1) heralding the initial state with a preceding measurement outcome (used for 
% $|\Phi_\text{ext}/\Phi_0 - 0.5| \leq 0.25$
$|\delta\Phi_\text{ext}| < 0.25\Phi_0$
)~\cite{Ding2023HighFidelity}, and (2) using a $\pi$-pulse to prepare the $\ket{1}$ state (used for 
% $|\Phi_\text{ext}/\Phi_0 - 0.5| > 0.25$)
$|\delta\Phi_\text{ext}| > 0.25\Phi_0$
~\cite{Blais2004}. 
We collected data over three separate cooldowns and found consistent trends across thermal cycles. %with identical filtering and attenuation configurations.
Flux biases were sampled more finely around the region near the degeneracy point, which serves as a typical operating point for gate operations~\cite{Ding2023HighFidelity, Rower2024CounterRotating}.
We started by calibrating readout as a function of flux bias, 
where at each $\Phi_\text{ext}$, we optimized readout frequency, recorded single-shots, and assigned IQ coordinates to the $\ket{0}$ and $\ket{1}$ states. 
A pulse length of $8~\mathrm{\mu s}$ was used for the initial characterization and temperature dependence datasets, while lengths of $8$-$10~\mathrm{\mu s}$ were used for the magnetic field sweep to account for diminishing readout contrast at higher fields.
In the flux bias range $|\delta\Phi_\text{ext}| \leq 0.25$, we measured $T_1$ by repeatedly sending readout pulses separated by a variable wait time $\tau$ (sequence shown in Fig.~\ref{supfig:T1-sequence}a), and then postselecting for the choice of initial state (either $\ket{0}$ or $\ket{1}$)~\cite{Ding2023HighFidelity}.
This avoided the need to calibrate a $\pi$-pulse at each flux bias.
Averaging over the qubit state single-shot measurements yielded a measure of the excited state probability $p_1$.
Delay times were sampled logarithmically to accomodate the large range of decay times across different flux biases.
At each working point, the qubit lifetime $T_1 = 1/\Gamma_1$ was extracted by fitting the decay to an exponential function $p(t) = A \exp(-t/T_1) + C$ (Fig.~\ref{supfig:T1-sequence}b).
Error bars correspond to one standard deviation error on the $T_1$ parameter. 
We excluded fits with a relative error in any parameter $\sigma_{s_i}/s_i> 0.3$, with $s_i\in\{A, T_1, C\}$.

While our dispersive readout scheme is ideally quantum non-demolishing (QND), in practice the readout pulse can induce transitions between the qubit states.
In our experiment, obtaining practical IQ-space separation of the $\ket{0}$ and $\ket{1}$ readout signals required long readout pulse durations and high power,
potentially giving rise to measurement-induced state transitions (MIST). Understanding and mitigating MIST is an active area of research~\cite{sank2016measurement, khezri2023measurement, dumas2024measurement}.
In our measurement scheme, since the readout pulse is not present during the free decay time, MIST should not affect the extracted decay rate. However, MIST can affect the measured steady-state populations $p_{0/1}$. Consequently, estimating $T_\text{eff}$ from steady-state population measurements should be done with care.
We measured in the limit of low-power, long-duration readout pulses, first confirming the populations were independent of readout power.
Then, we fit the resulting single-shot IQ data to the sum of two 2D Gaussians corresponding to the $\ket{0}$ and $\ket{1}$ states, from which we extracted the relative population amplitudes to obtain $p_0$ and $p_1$.
Using this method, we estimate on average $T_\text{eff} \approx 50~\mathrm{mK}$ in the flux range $|\delta\Phi_\text{ext}| \leq 15~\mathrm{m \Phi_0}$, which we use in Figs.~\ref{fig:t1-flux-baseline} and~\ref{fig:T1-B-sweep}.

We now comment on a benefit of MIST: 
as was also pointed out in Ref.~\cite{Somoroff2023}, the readout-induced transitions in our system enabled us to measure $T_1$ without $\pi$-pulse initialization up to $|\delta\Phi_\text{ext}| \approx 0.25$, or $f_{01} \approx 4.9~\mathrm{GHz}$. Although the thermal excited-state population at this frequency is only $0.9\%$, which would make postselecting on the $\ket{1}$ state impractical, there is a large enough excited-state population (which we attribute to MIST) that we can herald on the excited state even at high frequencies. 
Confirming this theoretically is beyond the scope of this work and left for future study.
However, we confirm that measurements both with and without $\pi$-pulse initialization yield consistent results. 

\section{Low-frequency flux noise} \label{app:low-freq-flux-noise}

\begin{figure}[t!]
    \centering
    \includegraphics[width=\columnwidth]{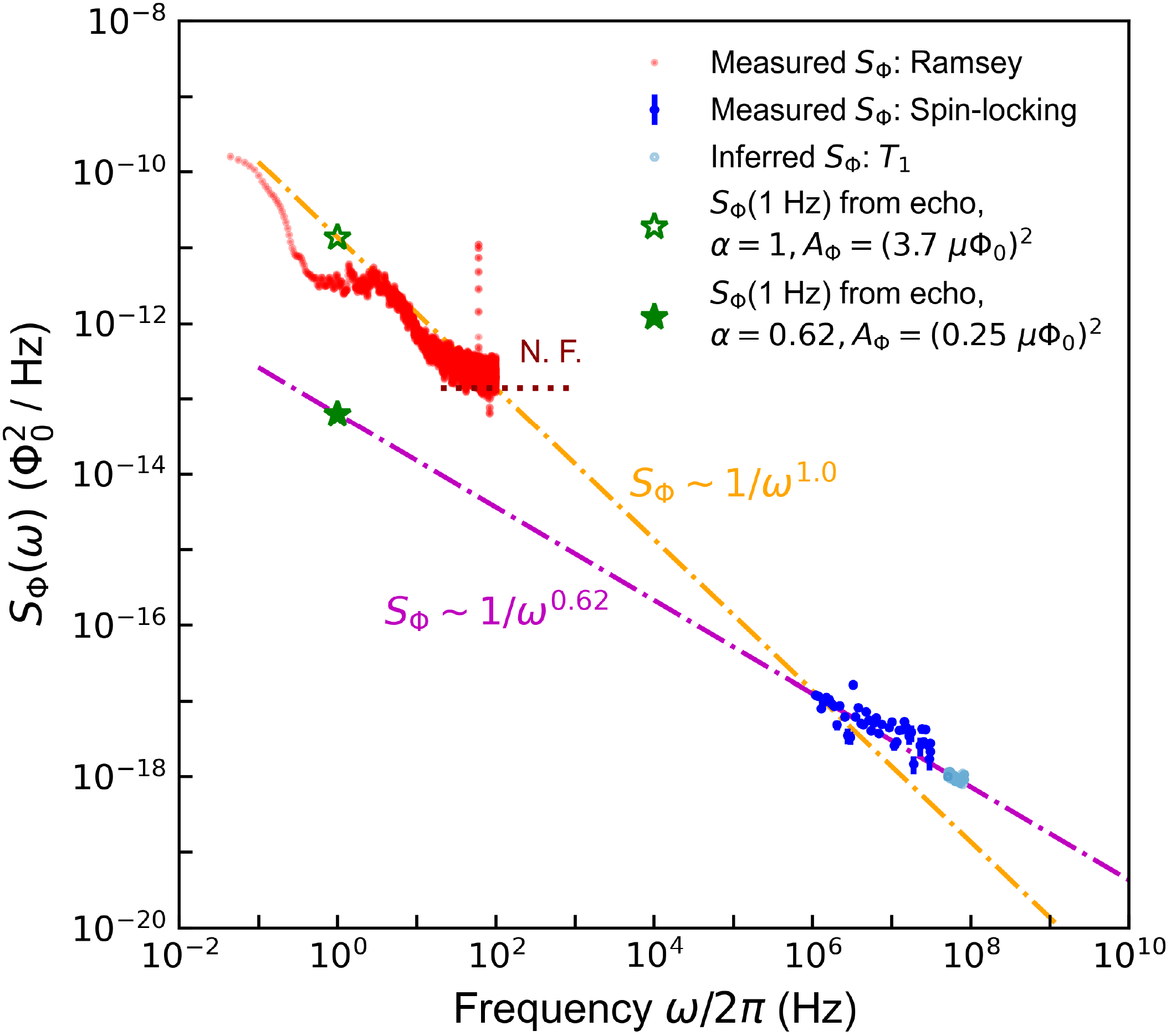}
    \caption{Flux noise spectrum at $B = 0~\mathrm{G}$. Data at $f< 100~\mathrm{Hz}$ were taken using single-shot Ramsey spectroscopy~\cite{yanSpectroscopyLowfrequencyNoise2012}. The peak in dephasing at 60 Hz corresponds to electronic noise, and the dotted line marks the measurement noise floor. Data in the range $1~\mathrm{MHz} \lesssim f \lesssim 30~\mathrm{MHz}$ are measured with spin-locking spectroscopy~\cite{yanRotatingframeRelaxationNoise2013, rower2023evolution, Sung2021Multi}. In the range $52~\mathrm{MHz} < f < 80~\mathrm{MHz}$, we plot the inferred flux noise spectrum from the $T_1$ measurements in Fig.~\ref{fig:t1-flux-baseline}. Green star markers indicate the noise magnitude $A_\Phi$ extracted from echo dephasing measurements, assuming $\alpha = 1$ (open marker) and $\alpha = 0.62$ (closed marker). Guides for $1/f^1$ and $1/f^{0.62}$ are shown. }
    \label{supfig:flux-noise-spectrum}
\end{figure}

% Temperature dependence figure here
\begin{figure*}[ht!]
    \centering
    \includegraphics[width=\textwidth]{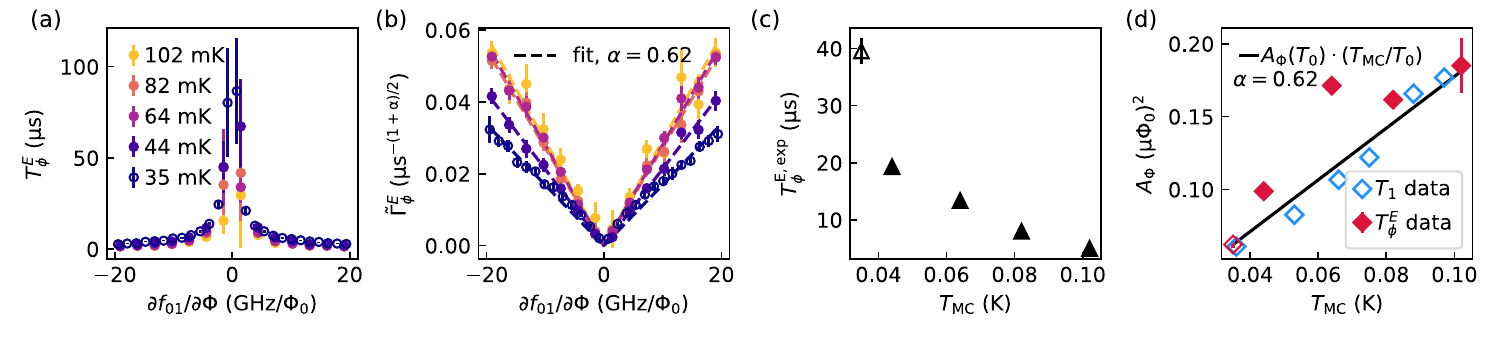}%
    \caption{Temperature dependence of spin-echo dephasing. 
    (a) Echo pure dephasing time $T_\phi^E$ versus flux sensitivity obtained from fits of decay traces to $A\exp[-t/(2T_1) - t/T_\phi^\text{E, exp}  - (t/T_\phi^E)^{1+\alpha}] + C$ with $\alpha = 0.62$ (see text).
    (b) $\tilde{\Gamma}_\phi^E \equiv 1/(T_\phi^E)^{(1+\alpha)/2}$ versus flux sensitivity as a function of temperature, along with fits to $\tilde{\Gamma}_\phi^E = |\partial \omega_{01}/\partial \Phi| \sqrt{A_\Phi z(\alpha)}$, where $z(\alpha) \equiv (-2 + 2^\alpha) \Gamma(-1-\alpha) \sin\left( \frac{\pi\alpha}{2} \right)$ (see Eq.~\ref{eq:coherence-arb-alpha}). $T_\text{MC} = 35~\mathrm{mK}$ data was taken during a different cooldown, indicated by open markers. 
    (c) Timescale of exponential component, $T_\phi^\text{E,exp}$, at each temperature. First value at $T_\text{MC} = 35~\mathrm{mK}$ is from measured echo dephasing at half-flux; remaining values are obtained from numerical fits to data in panel (b) (see text). 
    (d) Fitted flux noise magnitude $A_\Phi$ as a function of temperature, with red (blue) markers indicating values extracted from echo dephasing in panel (b) ($T_1$ decay in Fig.~\ref{fig:t1-flux-baseline}). A linear dependence $A_\Phi \propto T_\text{MC}$ is overlaid on the data. 
    }
    \label{supfig:flux-noise-temp-dependence}
\end{figure*}

Here we report on the characterization of low-frequency flux noise in our system. 
First, we characterize the quasi-static flux noise power spectral density (PSD) with single-shot Ramsey measurements~\cite{yanSpectroscopyLowfrequencyNoise2012, rower2023evolution} (Fig.~\ref{supfig:flux-noise-spectrum}). 
Notably, below $\sim$100 Hz we observe a distinct double-Lorentzian-like line shape  reminiscent of the Lorentzian-like features seen in Ref.~\cite{rower2023evolution} with application of an in-plane magnetic field and Ref.~\cite{yanRotatingframeRelaxationNoise2013} at zero magnetic field.
In contrast to Ref.~\cite{rower2023evolution}, these features appeared consistently at zero applied field and had consistent amplitudes and frequencies across multiple cooldowns. Under weak applied fields ($B \lesssim 10~\mathrm{G}$), the noise increased and the Lorentzian shoulders appeared to shift to higher frequencies (not shown). However, the increased dephasing noise at higher fields prevented us from further exploring the field dependence of the Ramsey PSD.

At higher frequencies, we measure the flux noise spectrum using spin-locking spectroscopy~\cite{yanRotatingframeRelaxationNoise2013, yanFluxQubitRevisited2016, rower2023evolution}, finding it consistent with a $1/f$-like dependence (Fig.~\ref{supfig:flux-noise-spectrum}).
We extract $S_\Phi(\omega)$ from measurements of the driven relaxation time $T_{1\rho} = 1/\Gamma_{1\rho}$, where $\Gamma_{1\rho} = \Gamma_1/2 + \Gamma_\nu$, using the conversion $S_\Phi(\omega) = S_{\omega_{01}}(\omega) \cdot (\partial \omega_{01}/\partial\Phi)^{-2} = 2\Gamma_\nu \cdot (\partial \omega_{01}/\partial\Phi)^{-2}$ ($\Gamma_\nu$ being the rate associated with Rabi frequency noise~\cite{yanRotatingframeRelaxationNoise2013}). 
These measurements were performed at qubit frequency $f_{01} \approx 204~\mathrm{MHz}$ ($|\partial f_{01} /\partial \Phi_0| \approx 19~\mathrm{GHz/\Phi_0}$).
Plotting these points alongside the flux noise at higher frequencies---in the range 52--80 MHz, inferred from the $T_1$ measurements in the main text---we find them consistent with a noise exponent $\alpha = 0.62$ and magnitude $A_\Phi = (0.25~\mathrm{\mu \Phi_0})^2$. This magnitude is lower than the measured value of $S_\Phi(1~\mathrm{Hz})$, indicating that flux noise does not follow the same frequency dependence throughout the frequency spectrum. We note that longitudinally-coupled charge noise is expected to contribute negligibly to the spin-locking decay rate given our junction array parameters~\cite{Randeria2024CQPS}. The apparent source of additional noise at $\sim$10--100 MHz is not known.

We further confirm the flux noise magnitude by measuring the spin-echo pure-dephasing rate $\Gamma_\phi^E$ as a function of flux sensitivity $|\partial f_{01} / \partial \Phi|$, which is also sensitive to $\sim$MHz frequencies ($T_\text{MC} = 35~\mathrm{mK}$ data in Fig.~\ref{supfig:flux-noise-temp-dependence}). By accounting for the exponent $\alpha = 0.62$---detailed below---we obtain the same value for $A_\Phi$ of $(0.25~\mathrm{\mu \Phi_0})^2$. 
Assuming $\alpha = 1$ for the same data yields $A_\Phi = (3.7~\mathrm{\mu \Phi_0})^2$, which is consistent with the typically observed magnitude for intrinsic flux noise~\cite{Randeria2024CQPS, Ardati2024Bifluxon, rower2023evolution, braumullerCharacterizingOptimizingQubit2020, yanFluxQubitRevisited2016, yanRotatingframeRelaxationNoise2013, yanSpectroscopyLowfrequencyNoise2012, slichter2012measurement, bylanderNoiseSpectroscopyDynamical2011}. We plot this value in Fig.~\ref{supfig:flux-noise-spectrum} for comparison.

\subsection{Accounting for $\alpha \neq 1$ in extracting $A_\Phi$ from echo dephasing measurements.}
% \todo{Fill this in. }
In the two-level approximation, the qubit is described by the Hamiltonian $\hat{H} = \hbar(\omega_{01} + \delta\omega(t))\sigma_z/2$, where $\omega_{01}$ is the qubit transition frequency and $\delta\omega(t)$ describes the stochastic frequency fluctuations induced by flux noise. A superposition state evolves with the phase $\phi(t) = \omega_{01}t + \delta\phi(t)$, where $\delta\phi(t) = \int_0^t dt' \delta\omega(t')$ is the stochastic phase fluctuation.
Averaging over many realizations of the fluctuating phase---assuming it obeys Gaussian statistics---leads to a qubit dephasing function $\langle e^{i\delta\phi}\rangle = e^{-\langle\delta\phi^2\rangle/2} \equiv e^{-\chi(t)}$. The coherence function $\chi(t)$ is found by integrating the overlap of the flux noise spectrum with the filter function corresponding to the pulse sequence, which for spin-echo dephasing is $g_E(\omega,t) = \sin^2(\omega t / 4) / (\omega t / 4)$~\cite{bylanderNoiseSpectroscopyDynamical2011}:
\begin{equation}
   \chi(t) = t^2 \left(\frac{\partial \omega_{01}}{\partial\Phi}\right)^2 \int_0^\infty d\omega S_\Phi(\omega) g_E(\omega,t),
\end{equation}
Taking the flux noise spectrum to be $S_\Phi(\omega) = A_\Phi(2\pi/\omega)^\alpha$, this integral evaluates to
\begin{align}
    \chi(t) &= t^{1+\alpha} \left(\frac{\partial \omega_{01}}{\partial\Phi}\right)^2 A_\Phi \times \notag \\ & (-2^{1-\alpha}) (-2 + 2^\alpha) \Gamma(-1-\alpha) \sin\left( \frac{\pi\alpha}{2} \right), \notag \\
    % & \equiv t^{1+\alpha} \left(\frac{\partial \omega_{01}}{\partial\Phi}\right)^2 A_\Phi \times  z(\alpha) 
    &\text{if } 0 < \alpha < 3 \label{eq:coherence-arb-alpha}
\end{align}
where $\Gamma(\cdot)$ is the gamma function, and which taken in the limit of $\alpha \rightarrow 1$ results in $\chi(t) = t^2 (\partial\omega_{01}/\partial\Phi)^2 A_\Phi \ln2$, the familiar Gaussian decay profile for $1/f$ flux noise~\cite{braumullerCharacterizingOptimizingQubit2020}.

We define $\chi(t) \equiv (t/T_\phi^E)^{1+\alpha} \equiv (\tilde{\Gamma}_\phi^E)^2 t^{1+\alpha}$ such that $\tilde{\Gamma}_\phi^E$ is proportional to $|\partial \omega_{01} / \partial \Phi|$ for arbitrary $\alpha \in (0,3)$ (reducing to $\tilde{\Gamma}_\phi^E = 1/T_\phi^E$ when $\alpha = 1$). 
We fit the spin-echo decay traces to $A\exp[-t/(2T_1) - t/T_\phi^\text{E, exp}  - (t/T_\phi^E)^{1+\alpha}] + C$, where $T_1$ is obtained from an immediately preceding relaxation measurement and $T_\phi^\text{E, exp}$ is estimated by measuring the pure-dephasing time at the degeneracy point, where the decay is exponential.
The remaining flux-noise-dephasing component $\tilde{\Gamma}_\phi^E$ is fit to a line against $|\partial \omega_{01} / \partial \Phi|$ to extract an estimate of $A_\Phi$.

\section{Temperature dependence of spin-echo dephasing} \label{app:T-dependent-dephasing}
% \begin{figure*}[th!]
%     \centering
%     \includegraphics[width=\textwidth]{supp_figs/supp_echo_vs_temperature_alpha_0_0617_wide_20250522.pdf}%{supp_figs/supp_fluxnoise_temp_dependence_Aphi_nbar_updated_20250224.pdf}
%     \caption{Temperature dependence of spin-echo dephasing. (a) Gaussian pure dephasing rate vs. flux sensitivity as a function of temperature, along with fit to $\Gamma_\phi^E = |\partial \omega_{01}/\partial \Phi| \sqrt{A_\Phi \ln2}$ (dashed lines). Decay traces were fit to $\exp[-\Gamma_1 t/2 - \Gamma_\phi^\mathrm{E,exp}t - (\Gamma_\phi t)^2]$, with $\Gamma_\phi^\mathrm{E,exp}$ extracted and indicated by dotted lines. 102, 64, and 44 mK data were taken in a single downwards sweep, while the 82 mK data was taken when sweeping back up; the 35 mK data was taken during a different cooldown.  (b) Fitted flux noise amplitude as a function of temperature, with effective $Q_L$ overlaid for comparison. (c) Effective average resonator photon number $\bar{n}$, estimated assuming $\Gamma_\phi^\mathrm{E,exp}$ is purely due to photon shot noise. Inset: effective resonator temperature for $\bar{n}$ in the main plot.}
%     \label{supfig:flux-noise-temp-dependence}
% \end{figure*}
% 
To further examine the temperature dependence of $T_1$ at low frequencies and confirm whether it is consistent with being limited by flux noise, we performed a separate temperature sweep in which we measured the spin-echo pure-dephasing time as a function of the flux noise susceptibility (Fig.~\ref{supfig:flux-noise-temp-dependence}a).
We stress that the $T_\text{MC} = 35~\mathrm{mK}$ measurements are taken in a separate cooldown from the $T_\text{MC} > 35~\mathrm{mK}$ sweep, but we include them in the same plot for completeness.
At each value of $T_\text{MC}$, we repeated the procedure detailed in the previous section to extract $A_\Phi$ versus temperature, assuming a fixed exponent $\alpha = 0.62$ obtained at base temperature (Fig.~\ref{supfig:flux-noise-temp-dependence}c). (We acknowledge that this exponent may change with temperature~\cite{antonMagneticFluxNoise2013}, which we leave for a future study.) For temperatures $T_\text{MC} > 35~\mathrm{mK}$, for which we did not have a separate dephasing measurement at half-flux, we numerically find the value of $T_\phi^\text{E, exp}$ such that the remaining decay contribution scales most linearly with the flux noise susceptibility (Fig.~\ref{supfig:flux-noise-temp-dependence}b). Explicitly, we obtain this by maximizing the coefficient of determination $R^2$ for the fit of $\tilde{\Gamma}_\phi^E$ vs. $|\partial f_{01}/\partial\Phi|$ to a line.
We plot the extracted $A_\Phi$ as a function of $T_\text{MC}$ in Fig.~\ref{supfig:flux-noise-temp-dependence}d, on which we also overlay the values inferred from fitting the temperature-dependent $T_1$ data in Fig.~\ref{fig:T-sweep}. We plot a phenomenological linear dependence $A_\Phi(T) = A_\Phi(T_0) \cdot (T/T_0)$, which approximately aligns with the data. 
The coincidence of the spin-echo and relaxation trends with temperature suggests that they both probe the same underlying noise source.

\section{Models for $T_1$ decay} \label{app:models}
Here we discuss the energy relaxation models considered in this study.
For each loss mechanism, we consider a fluctuating noise source $\lambda(t)$ coupled to a qubit operator $\hat{D}_\lambda$ through the interaction Hamiltonian $\hat{H}_\text{int} = \hat{D}_\lambda \lambda(t)$. This results in transitions between the qubit states through energy exchange with the environment, where Fermi's golden rule gives the transition rates for excitation ($\ket{0}\rightarrow \ket{1}$) and relaxation ($\ket{1} \rightarrow \ket{0}$),
\begin{equation}
    \Gamma_{\uparrow / \downarrow} = \frac{1}{\hbar^2} |\bra{0} \hat{D}_\lambda \ket{1} |^2 \tilde{S}_\lambda(\mp \omega_{01}),
\end{equation}
where the two satisfy detailed balance in thermal equilibrium, $\Gamma_\uparrow/\Gamma_\downarrow = \exp(-\hbar{\omega_{01}}/k_BT_\text{eff})$.
Here, $\tilde{S}_\lambda(\omega) = \int_{-\infty}^\infty  \langle \lambda(0) \lambda(t)\rangle e^{-i\omega t}dt $ is the Fourier transform of the unsymmetrized autocorrelation function of $\lambda(t)$ and denotes the bilateral noise spectral density, with positive (negative) values of $\omega$ corresponding to the emission of energy to (absorption of energy from) the environment. 
The depolarization rate is given by the sum of the two rates, $\Gamma_1 = 1/T_1 =  \Gamma_\uparrow + \Gamma_\downarrow$, and is therefore sensitive to the symmetrized (classical) noise spectrum,
\begin{equation}
    S_\lambda(\omega) = \frac{\tilde{S}_\lambda(\omega) + \tilde{S}_\lambda(-\omega)}{2}.
\end{equation}
For dissipation arising from a combination of mechanisms, we sum their contributions to get Eq.~\ref{eq:gamma1} in the main text,
\begin{equation}
    \Gamma_1 = \sum_\lambda \frac{2}{\hbar^2} |\bra{0}{\hat{D}_\lambda}\ket{1}|^2 S_\lambda(\omega_{01}).
\end{equation}
For circuit-model noise, we follow
the Caldeira-Leggett model, which describes the voltage fluctuations across the circuit in terms of the real part of its impedance $Z(\omega)$~\cite{devoret1995quantum,  clerkIntroductionQuantumNoise2010}:
\begin{equation}
    \tilde{S}_\text{V} (\omega) = \frac{2 \text{Re}[Z(\omega)] \hbar\omega}{1-e^{-\hbar\omega/k_BT}}.
\end{equation}
This is the familiar two-sided Johnson-Nyquist noise power spectrum~\cite{clerkIntroductionQuantumNoise2010}, which reduces to Nyquist or Ohmic noise at large positive frequencies ($\hbar\omega \gg k_B T$) and thermal or Johnson noise at low frequencies ($|\hbar\omega| \ll k_B T$).
The symmetrized voltage spectral density is given by 
% 
% \begin{equation}
%     S_\text{V}^+ (\omega) = 2 \text{Re}[Z(\omega)] \hbar\omega \coth\left( \frac{\hbar\omega}{2 k_B T} \right).
% \end{equation}
\begin{equation}
    S_\text{V} (\omega) = \text{Re}[Z(\omega)] \hbar\omega \coth\left( \frac{\hbar\omega}{2 k_B T} \right).
\end{equation}
We use this treatment (and its equivalent form in terms of current noise and circuit admittance) to model inductive loss, dielectric loss, QP tunneling, Purcell, and radiation to the control lines, as described below. 

\subsection{$1/f$ flux noise}

Low-frequency magnetic flux noise couples to the qubit phase operator through the inductive term of the Hamiltonian, 
$\lambda(t) = \delta\Phi_\text{ext}(t)$ and $\hat{D}_\lambda = (2\pi E_L/\Phi_0) \hat{\phi}$,
% $H_\text{int} = E_L \hat{\phi} \phi_\text{ext}$, 
yielding a depolarization rate
% \begin{equation}
%     \Gamma_1^{1/f} = \left(\frac{2\pi E_L}{\hbar \Phi_0}\right)^2 |\langle 0|\hat{\phi}|1 \rangle |^2 S_{\Phi}^+(\omega_{01}), 
% \end{equation}
\begin{equation}
    \Gamma_1^{1/f} = 8 \left(\frac{\pi E_L}{\hbar \Phi_0}\right)^2 |\bra{0} \hat{\phi} \ket{1}|^2 S_{\Phi}(\omega_{01}), 
\end{equation}
where the flux noise power spectrum is defined as~\cite{yanSpectroscopyLowfrequencyNoise2012, yanRotatingframeRelaxationNoise2013, yanFluxQubitRevisited2016}
\begin{equation}
    S_\Phi(\omega) = A_\Phi \left(\frac{2\pi}{\omega} \right)^\alpha
\end{equation}
% where we define
% \begin{equation}
%     \frac{1}{2} S_\Phi^+(\omega) = \frac{A_\Phi}{(\omega/2\pi)^\alpha}
% \end{equation}
As mentioned, we obtain the $1/f$ noise amplitude by analyzing spin-echo dephasing and spin-locking data (\ref{app:low-freq-flux-noise}). In our qubit, this is believed to be the dominant loss mechanism at half-flux, where the qubit frequency is $f_{01} = 52~\mathrm{MHz}$. 

\subsection{Inductive loss}
Inductive loss is related to current fluctuations in the inductor (implemented here by a Josephson-junction array superinductor) leading to bias-flux fluctuations, $\lambda(t) = \delta\Phi_\text{ext}(t)$ and $\hat{D}_\lambda = (2\pi E_L/\Phi_0) \hat{\phi}$, with the power spectrum given by Johnson-Nyquist current noise:
\begin{equation}
    S_{\Phi,\mathrm{ind}}(\omega) = \frac{\hbar\Phi_0^2}{4\pi^2 E_L Q_L} \coth\left(\frac{\hbar\omega}{2k_BT_\mathrm{eff}} \right), \label{eq:ind-loss-psd}
\end{equation}
where $Q_L$ is the inductive quality factor. 
This results in a depolarization rate
\begin{equation}
    \Gamma_1^\text{ind} = \frac{2 E_L}{\hbar Q_L} |\langle 0|\hat{\phi}|1 \rangle |^2 \coth \left( \frac{\hbar\omega_{01}}{2k_B T_\text{eff}} \right).
\end{equation}
An example source of such noise is QP tunneling in the junction array~(\ref{subsec:qp-array}), but here we discuss generic loss through the inductor as the dual of capacitive loss~(\ref{subsec:diel-loss}). 
This treatment has been used to model the coherence of nanowire fluxonium qubits~\cite{Hazard2019Nanowire}, though the microscopic origin of such generic inductive noise is unclear.

As the low-frequency behavior is challenging to distinguish from $1/f$ flux noise with frequency dependence alone ($\coth(\hbar\omega/2k_BT) \approx 2k_BT/\hbar\omega$), we can derive an approximate relation between the flux noise amplitude and inductive quality factor:
\begin{equation}
    Q_L^{-1} \approx \frac{4\pi^3 E_L}{k_BT\Phi_0^2} \left(\frac{\omega}{2\pi} \right)^{1-\alpha} A_\Phi.
    \label{eq:eff-QL}
\end{equation}
For our data (Fig.~\ref{fig:t1-flux-baseline}a), $Q_L = 3.2\times10^8$ equivalently describes our data with $T_\text{eff} = 50~\mathrm{mK}$.

\subsection{Dielectric loss}\label{subsec:diel-loss}
Dissipation through lossy dielectric materials in the device is modeled as Johnson-Nyquist voltage noise across the capacitor, or equivalent noise in the gate charge $\lambda(t) = \delta Q_g(t) = 2en_g(t)$, with $\hat{D}_\lambda = (4E_C/e)\hat{n}$. 
% with a coupling Hamiltonian $H_\text{int} = 8 E_C \hat{n} n_g$.
The loss rate can be written in terms of the capacitive loss tangent $\tan\delta_C(\omega)$ as
\begin{equation}
    \Gamma_1^\text{diel} = \frac{16E_C}{\hbar} |\langle 0|\hat{n}|1 \rangle |^2 \tan \delta_C(\omega_{01}) \coth \left( \frac{\hbar\omega_{01}}{2k_B T_\text{eff}} \right).
\end{equation}
Although this form directly reflects the electric-dipole nature of the coupling, it is also instructive to write it in terms of the phase matrix element, using the relation $\matel{n} = (\hbar\omega/8E_C)\matel{\phi}$~\cite{Zhang2021Universal}, as
\begin{equation}
    \Gamma_1^\text{diel} = \frac{\hbar\omega_{01}^2}{4E_C} |\langle 0|\hat{\phi}|1 \rangle|^2 \tan \delta_C(\omega_{01}) \coth \left( \frac{\hbar\omega_{01}}{2k_B T_\text{eff}} \right),
\end{equation}
which offers direct comparison to the expression for flux noise.
The frequency dependence of the loss tangent is empirically described by a power law: $\tan\delta_C(\omega)=\tan\delta_C^0 (\omega/\omega_\text{ref})^\epsilon$, where $\omega_\text{ref} = 2\pi\times6~\mathrm{GHz}$~\cite{nguyenHighCoherenceFluxoniumQubit2019}. 

We find that for our base temperature $T_1$ data, direct application of this model to the computational levels cannot simultaneously describe the data across all flux biases. 
As one plausible explanation for this discrepancy, we address the effect of heating to levels outside the computational subspace, described in \appref{app:heating}.

\subsection{QPs at the small junction}
For current noise from QPs tunneling across the small junction, we have $\lambda(t) = \delta I_\text{QP}(t)$ and $\hat{D}_\lambda = (\Phi_0/\pi) \sin (\hat{\phi}/2)$, which results in a decay rate~\cite{smith2019design, Pop2014Coherent, Catelani2011Relaxation}
% \begin{equation}
%     \Gamma_1^\text{QP}=\frac{2\omega_{01}\phi_0}{\hbar} |\langle 0 | \sin\frac{\hat{\phi}}{2} | 1 \rangle|^2 \text{Re}[Y_\text{QP}(\omega_{01})] \left( 1 + \coth\frac{\hbar\omega}{2k_B T_\text{eff}}\right),
% \end{equation}
% \begin{equation}
%     \Gamma_1^\text{QP}=\frac{2\phi_0}{\hbar^2} |\langle 0 | \sin\frac{\hat{\phi}}{2} | 1 \rangle|^2 S_\text{QP}^+(\omega_{01}),
% \end{equation}
\begin{equation}
    \Gamma_1^\text{QP}= 2\left(\frac{\Phi_0}{\hbar\pi}\right)^2 |\langle 0 | \sin(\hat{\phi}/2) | 1 \rangle|^2 S_\text{QP}(\omega_{01}),
\end{equation}
where the spectral density 
$S_\text{QP}(\omega)=\hbar\omega \text{Re}[Y_\text{QP}(\omega)] \coth\left(\hbar\omega/2k_B T\right)$ 
% $S_\text{QP}^+(\omega)=2\hbar\omega \text{Re}[Y_{qp}(\omega)] \coth\left(\hbar\omega/2k_B T\right)$ 
describes Johnson-Nyquist current noise related to the dissipative part of the admittance~\cite{smith2019design},
\begin{align}
    \text{Re}[Y_\text{QP}(\omega)] &= \sqrt{\frac{2}{\pi}} \frac{8E_J}{R_K\Delta} \left( \frac{2\Delta}{\hbar\omega} \right)^{3/2} x_\text{QP} \times \\
    &\sqrt{\frac{\hbar\omega}{2k_B T}} K_0 \left(\frac{\hbar|\omega|}{2k_B T}\right) \sinh\frac{\hbar\omega}{2k_B T}.
\end{align}
Here, $\Delta$ is the Al superconducting gap, $R_K=h/e^2$ the normal resistance quantum, $K_0$ a modified bessel function of the second kind, and $x_\text{QP}$ the relative QP density, which is related to the temperature by~\cite{serniak2018hot}
\begin{equation}
    x_\text{QP}=x_\text{QP}^0 + \sqrt{2\pi k_B T/\Delta} e^{-\Delta/k_B T}.
\end{equation}
The decay rate vanishes at the degeneracy point, resulting in theoretical protection from QP-tunneling-induced decoherence.

Although it is challenging in our experiment to estimate the contribution to $T_1$ from QPs in the small junction, we can begin by placing a bound assuming this mechanism limits the $T_1$ at integer flux bias, which gives $x_\text{QP} \approx 3\times 10^{-6}$. This is more than an order of magnitude higher than observed in prior experiments~\cite{Pop2014Coherent, serniak2018hot}, as well as the upper bound estimated from QP tunneling in the junction array (see below). 
Additionally, the temperature dependence of $x_\text{QP}$ below 100 mK is very weak, and is inconsistent with the measured data, even when including heating out of the computational subspace. 
We therefore conclude that QP tunneling in the small junction does not limit the $T_1$ across the flux range. 
In the main text we assume $x_\text{QP} \approx 1\times 10^{-7}$.

\subsection{QPs in the junction array}\label{subsec:qp-array}
QPs tunneling in the junction array are suspected to constitute a source of inductive loss in fluxonium qubits~\cite{Vool2014NonPoissonian}, with a relaxation rate described by
% \begin{equation}
%     \Gamma_1^\text{QPA} = E_L |\langle 0 | \frac{\hat{\phi}}{2}|1\rangle|^2 \tilde{S}_\text{QP}^+(\omega_{01}),
% \end{equation}
\begin{equation}
    \Gamma_1^\text{QPA} = 2\left(\frac{\Phi_0}{\hbar\pi}\right)^2 E_L  |\langle 0 | \frac{\hat{\phi}}{2}|1\rangle|^2 \bar{S}_\text{QP}(\omega_{01}), \label{eq:QPA}
\end{equation}
where $\bar{S}_\text{QP}(\omega) = S_\text{QP}(\omega) / E_J$~\cite{Catelani2011Relaxation}.
The $T_1$ at half-flux places an upper bound of $x_\text{QPA} \approx 7\times 10^{-9}$ on QPs in the array.
We note that there is some evidence for relative QP densities differing between the small individual junction and the array junctions~\cite{Vool2014NonPoissonian}.
Since this mechanism scales similarly with flux noise and inductive loss near half-flux, it is also indistinguishable from these mechanisms with flux dependence alone.
However, as we show in~\ref{app:half-flux-compare}, the temperature dependence of $T_1$ at half-flux is inconsistent with this model, so we neglect QPs in the array in the main analysis. 

\subsection{Radiation to the charge and flux lines}
Radiative decay to the charge (flux) lines is estimated by Johnson-Nyquist voltage (current) noise from a $50~\Omega$ environment connected to the qubit via a coupling capacitance (inductance) set by the control line geometry. 

\textbf{Charge line.} 
Voltage fluctuations on the charge-drive line $\lambda(t) = \delta V_d(t)$ couple to the qubit with $\hat{D}_\lambda = (4E_C C_d/e) \hat{n}$, where $C_d$ is the coupling capacitance.
From this we can compute the decay rate with Fermi’s golden rule,
\begin{equation}
    \Gamma_1^\text{ch} = 2\left(\frac{4E_C C_d}{e\hbar} \right)^2 |\langle 0 |\hat{n}|1\rangle|^2 S_{V}(\omega),
\end{equation}
where the symmetrized voltage spectral density 
% $S^+_{V} (\omega) = S_V(\omega) + S_V(-\omega)$ 
is the Johnson-Nyquist noise from a feedline with impedance $Z_0 = 50~\Omega$~\cite{clerkIntroductionQuantumNoise2010}: 
\begin{equation}
    S_{V}(\omega) = Z_0 \hbar \omega [2 n_B(\omega) + 1].
\end{equation}
Here, $n_B(\omega) = 1 / [\exp(\hbar \omega / k_BT) -1]$ is the Bose-Einstein occupation factor at temperature $T$. 
However, it is possible that the number of noise photons at the base-temperature stage of the DR is greater than the number predicted by thermal occupation because photons from higher stages may not be fully attenuated. To account for this, we compute the number of noise photons at stage $i$ of the DR given the attenuation $A_i$ at that stage and the number of photons coming from the prior attenuation stage $n_{i-1}$~\cite{krinner2019engineering}:
\begin{equation}
    n_i(\omega) = \frac{n_{i-1}(\omega)}{A_i} + \frac{A_i - 1}{A_i} n_B(T_{i, att}, \omega).
\end{equation}
This model treats each attenuator as a beamsplitter with a $1: A_i - 1$ transmission ratio, which lets through $1 / A_i$ of the signal photons and $(A_i-1)/A_i$ of the thermal photons at stage i. 
Using this model, we replace $n_B$ in the expression for  $S_{V} (\omega)$ with the the photon number at the MC stage $n_\text{MC}$ calculated for our setup (see~\ref{app:setup} for the attenuation configuration; we ignore the final-stage low-pass filter for simplicity, which passes radiation at the qubit frequency near half-flux but further attenuates higher-frequency radiation).
Using coupling capacitance $C_d = 20~\mathrm{aF}$ estimated from the design parameters, we estimate $T_1 \gtrsim 1\times 10^5~\mathrm{\mu s}$ at half-flux, and $T_1 \gtrsim 100~\mathrm{\mu s}$ at the plasmon transition, far above the measured data in both cases. For the plasmon transition, this would be a conservative lower bound because of the additional high-frequency attenuation provided by the final-stage low-pass filter.

\textbf{Flux line.} 
We do an analogous estimate for radiation to the flux-drive line, where current fluctuations from a $50~\mathrm{\Omega}$ environment $\lambda(t) = \delta I_d(t)$ couple to the phase operator with $\hat{D}_\lambda = (2\pi E_L M_d)/\Phi_0 \hat{\phi}$, where $M_d$ is the mutual inductance between the flux line and the qubit. The decay rate for radiation to the flux line is given by
\begin{equation}
    \Gamma_1^\text{fl} = 8 \left( \frac{\pi E_L M_d}{\hbar \Phi_0} \right)^2 |\langle 0 | \hat{\phi} |1\rangle |^2 S_I(\omega_{01})
\end{equation}
where  
\begin{equation}
    S_I(\omega) = \hbar\omega \text{Re}[Y(\omega)] [2 n_\text{B}(\omega) + 1],
\end{equation}
is the Johnson-Nyquist current noise spectrum in terms of the real part of the admittance, $\text{Re} [Y(\omega)] = Z_0/[Z_0^2 + \omega^2 (M_d^2/L)^2]$, and we again make the replacement $n_B(\omega) \rightarrow n_\text{MC}$.
In our qubit, the measured mutual inductance is $M_d = \Phi_0 / 21.5~\mathrm{mA}$, resulting in a $T_1$ limit of $\gtrsim 0.2~\mathrm{s}$ at the sweet spot, so we neglect it in our analysis. 

\subsection{Purcell decay through the readout resonator}

\begin{figure}[tb]
    \centering
    \includegraphics[width=\columnwidth]{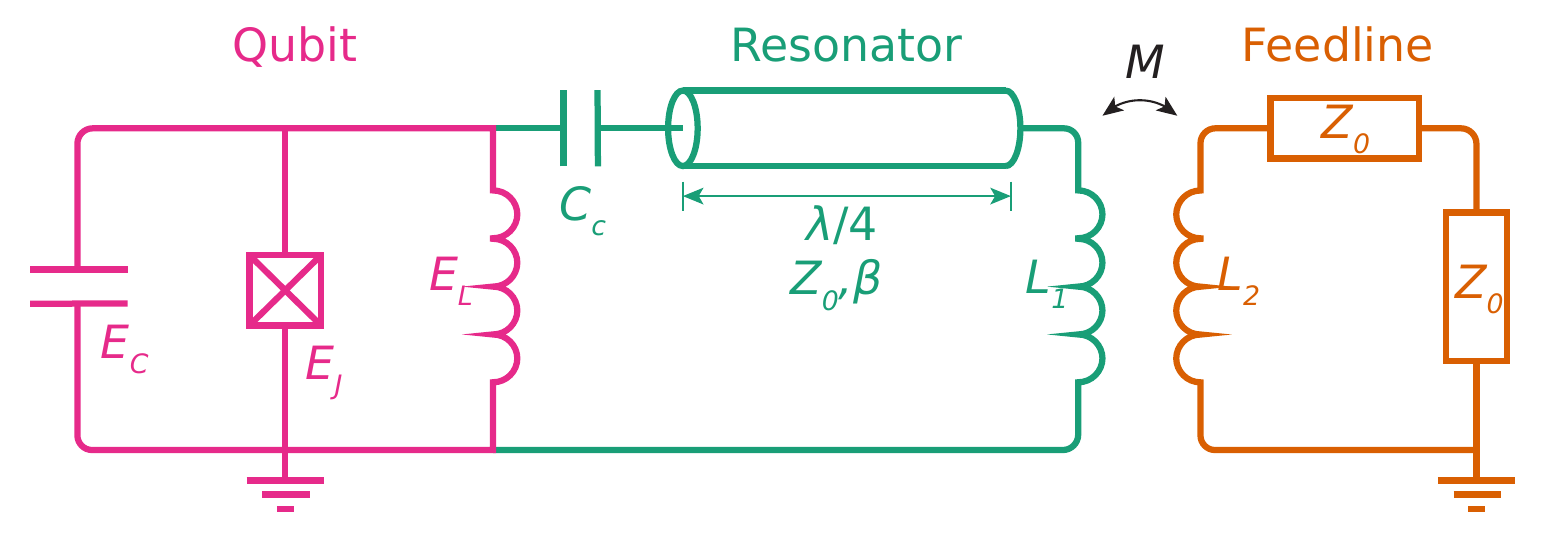}
    \caption{Equivalent circuit used to model Purcell decay.}
    \label{fig:purcell-circuit}
\end{figure}

We model the fluxonium’s Purcell decay rate through the readout resonator by considering the equivalent-circuit model of the coupled qubit-resonator system, shown in Fig.~\ref{fig:purcell-circuit}. A $\lambda/4$ transmission-line resonator with characteristic impedance $Z_0 = 50~\Omega$ is capacitively coupled to the qubit on one end, and inductively (over-)coupled to a feedline on the other end, also with $Z_0 = 50~\Omega$.
The corresponding Fermi’s golden rule expression for $\Gamma_1$ decay in the qubit is
\begin{equation}
    \Gamma_1^\text{Purcell} = \frac{8 e^2\omega}\hbar \left(\frac{C_c}{C_\Sigma}\right) |\langle 0|\hat{n} | 1 \rangle|^2 \text{Re}[Z_\text{in}(\omega)] \coth\left( \frac{\hbar \omega}{2k_BT_\text{res}} \right), \label{eq:gamma1-purcell}
\end{equation}
where $C_\Sigma = e^2/2E_C$, $C_c$ is the coupling capacitance between the qubit and the resonator, 
\begin{equation}
    C_c = \frac{\hbar g C_\Sigma}{2\omega_\text{res} e} \sqrt{\frac{\pi}{2\hbar Z_0}}, \label{eq:Cc-purcell}
\end{equation}
and $Z_\text{in}(\omega)$ is the impedance of the environment seen by the qubit, describing the filtering of the bath modes by the resonator.
For the fundamental mode at $\omega_\text{res}$, this is given by~\cite{pozar2021microwave}
\begin{equation}
    Z_\text{in}(\omega) = Z_0 \frac{\omega^2 M^2 \cot (\frac{\pi\omega}{2\omega_\text{res}}) + 2j Z_0^2}{2Z_0^2 \cot (\frac{\pi\omega}{2\omega_\text{res}}) + j\omega^2 M^2}, \label{eq:Zin-purcell}
\end{equation}
where $M$ is the mutual inductance between the resonator and feedline, and we have neglected the self inductances $L_1 = L_2 = 0$ (which only serve to shift $\omega_\text{res}$). 
We rewrite $M$ in terms of experimentally accessible parameters $\omega_\text{res}$ and $Q = \omega_\text{res}/\kappa$ by approximating the circuit near resonance with a parallel RLC resonator and assuming weak coupling to the feedline ($\omega M \ll Z_0$), giving 
\begin{equation}
    M = \frac{Z_0}{\omega_\text{res}} \sqrt{\frac{\pi}{2Q}}. \label{eq:M-purcell}
\end{equation}
Using Eqs.~\ref{eq:gamma1-purcell}-\ref{eq:M-purcell}, we calculate the Purcell lifetime with a resonator temperature $T_\text{res} = 70~\mathrm{mK}$, estimated from $T_\phi^E = 71~\mathrm{\mu s}$ at the sweet spot, assuming fully shot noise-limited dephasing. 
This gives a limit of $T_1^\text{Purcell} \approx 80~\mathrm{\mu s}$ at the plasmon transition, an order of magnitude higher than our measured $T_1$.

\section{Comparison of models at half-flux} \label{app:half-flux-compare}
\begin{figure}[t!]
    \centering
    \includegraphics[width=\columnwidth]{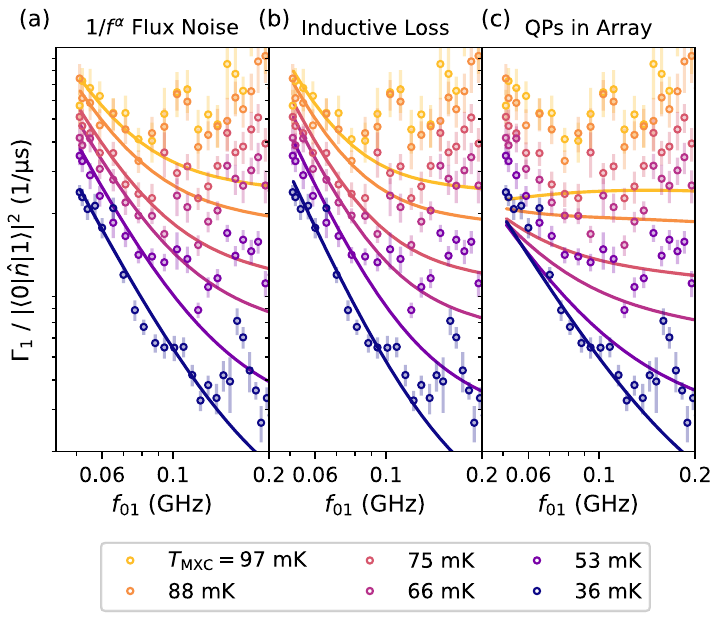}
    \caption{Comparison of temperature-dependence of models at  low-frequency (half-flux). (a) $1/f^\alpha$ flux noise, with $A_\Phi(T_0) = (0.22~\mathrm{\mu \Phi_0})^2$, $\alpha = 0.62$, and $A_\Phi(T) = A_\Phi(T_0) \cdot (T/T_0)$, $T_0 =36~\mathrm{mK}$. (b) Inductive loss with $Q_L=1.8\times10^8$. (c) QP tunneling in the junction array, with $x_\text{qpa}=7.4\times10^{-9}$. In each case, all other models are held fixed with the same parameter values as in the main text, and evaluated using $N$ levels. }
    \label{supfig:T-lowfreq-model-compare}
\end{figure}

Here we compare the temperature dependence of several phase-coupled loss mechanisms at half-flux: $1/f^\alpha$ flux noise with a phenomenological temperature dependence $A_\Phi(T) = A_\Phi(T_0) \cdot (T/T_0)$, inductive loss (Eq.~\ref{eq:ind-loss-psd}), and QP tunneling in the junction array (Eq.~\ref{eq:QPA}).
Fig.~\ref{supfig:T-lowfreq-model-compare} plots the temperature dependence of $\Gamma_1/\matel{n}^2$ (same data as in Fig.~\ref{fig:T-sweep}), highlighting the low-frequency region. Each panel shows a loss model comprising one of the three phase-coupled mechanisms, Purcell decay, dielectric loss, and QP tunneling across the small junction. The three common models share the same parameters as Fig.~\ref{fig:T-sweep}b.
While low-frequency noise measurements at a single temperature can be reasonably well-described by any one of these candidates, making it challenging to distinguish them with frequency dependence alone, it is apparent from the temperature dependence that $1/f^\alpha$ flux noise (Fig.~\ref{supfig:T-lowfreq-model-compare}a) and inductive loss (Fig.~\ref{supfig:T-lowfreq-model-compare}b) give the best agreement. We again stress that the microscopic model for inductive loss is unknown.
On the other hand, the poor agreement of QP tunneling in the array junctions (Fig.~\ref{supfig:T-lowfreq-model-compare}c) allows us to exclude this mechanism from our analysis. 

\section{Effect of heating outside the computational subspace} \label{app:heating}
Here we detail the decoherence model incorporating levels outside the computational subspace of the circuit. Although one might typically consider $T_1$ of only the $\ket{0}\leftrightarrow \ket{1}$ transition in design, the transition rates between higher energy levels can be significant for the effective (measured) depolarization, particularly at flux biases where the energy of the transition approaches the thermal energy scale. We use this model to partially explain the inconsistency of the measured data with the model of decoherence within the 2-level qubit subspace. 

\subsection{$N$-Level Decoherence Model}
\begin{figure}
    \centering
    \includegraphics[width=\columnwidth]{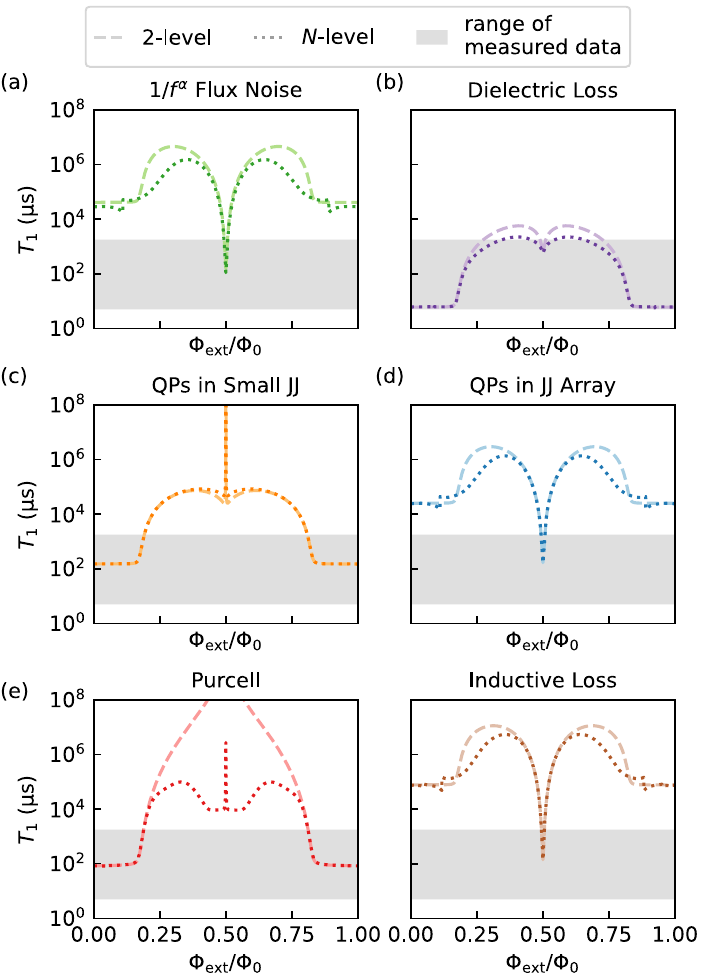}
    \caption{Comparison of 2-level vs. $N$-level models. Gray shaded region indicates range of measured $T_1$ data, showing that heating events impact only dielectric loss significantly in the relevant range.}
    \label{supfig:N-level-model-compare}
\end{figure}

Consider a vector $\vec{p}(t) = \begin{pmatrix} p_0 & p_1 & \cdots & p_{N-1} \end{pmatrix}^T$ with length $N$, representing the populations of the first $N$ levels of the fluxonium at time $t$. The dynamics of the level occupation probabilities $p_i$, are described by the matrix rate equation
$\partial_t\vec{p}(t) = B \vec{p}(t)$, where
\begin{equation}
    B = \begin{pmatrix} 
    -\sum\limits_i\Gamma_{0\rightarrow i} & \Gamma_{1\rightarrow 0}  & \cdots & \Gamma_{N\rightarrow 0}  \\ 
    \Gamma_{0\rightarrow 1}  &  - \sum\limits_i\Gamma_{1\rightarrow i} & \cdots & \vdots\\
    \vdots & \vdots & \ddots & \Gamma_{N\rightarrow N-1}    \\
    
    \Gamma_{0\rightarrow N}  & \Gamma_{1\rightarrow N}  & \cdots &  - \sum\limits_i \Gamma_{N\rightarrow i} 
    \end{pmatrix} \label{eq:B-matrix-app}
\end{equation}
is an $N\times N$ matrix, and $\Gamma_{i\rightarrow j}$ is calculated from Fermi's golden rule. We note that in this model, we assume the frequency dependence of each noise mechanism is known, and that the system is in thermal equilibrium with its environment, so the upwards and downwards transition rates satisfy detailed balance, $\Gamma_{i\rightarrow j} / \Gamma_{j\rightarrow i} = \exp[\hbar(\omega_i-\omega_j)/k_BT]$.
We can diagonalize the matrix, $B=VSV^{-1}$ where
\begin{equation}
    S = \begin{pmatrix} 0 & 0 & \cdots &\cdots & 0 \\
    0 & -\gamma_1 & 0 & \cdots & 0 \\
    0 & 0 & -\gamma_2 & \cdots & 0 \\
    \vdots & & & \ddots & \vdots \\
    0 & 0 & \cdots & \cdots & -\gamma_N
\end{pmatrix}
\end{equation}
contains the eigenvalues on the diagonal, and $V$ contains the eigenvectors $\vec{v}_i$ as columns. The eigenvalues $-\gamma_i$ directly correspond to the decay rates of each eigenvector of the system, with one of the eigenvalues $\gamma_0 = 0$ corresponding to the steady state vector. 
The solution is found by directly exponentiating the matrix: $\vec{p}(t) = e^{Bt}\vec{p}(0) = Ve^{St} V^{-1} \vec{p}(0)$. Each eigenvector evolves separately, 
\begin{equation}
    \vec{p}(t) = c_0 \vec{v}_0 + c_1 \vec{v}_1e^{-\gamma_1 t}  + \cdots + c_{N-1} \vec{v}_{N-1} e^{-\gamma_{N-1} t}, \label{eq:p-t-eigenvector-evolution}
\end{equation}
where $\begin{pmatrix} c_0 & c_1 & \cdots & c_{N-1} \end{pmatrix}^T = V^{-1}\vec{p}(0) \equiv \tilde{\vec{p}}(0) $ are the coefficients of the initial vector in the eigenbasis of $B$. We define the eigenvectors such that they have unit length, $||\vec{v}_i||=1$. 
This gives us a way to define the effective $\ket{0}\leftrightarrow \ket{1}$ decay rate in an $N$-level system: $\Gamma_1^\text{eff} = \gamma_{k_\text{max}}$, where $k_\text{max} = \text{argmax}_{i\neq0}|c_i|^2$. In other words, it is the eigenvalue of the non-steady-state normal mode sharing the largest overlap with the initial state of the system. Here we initialize the population entirely in the $\ket{1}$ state, $p_1=1,~p_j=0$ for $j\neq 1$~\footnote{We note that for initialization of up to 98\% in $\ket{0}$, the $\Gamma_1^\text{eff}$ is identical at nearly all flux biases given our parameters. At some isolated biases, significant overlap is contained in multiple eigenvectors and a different eigenvector may be selected. In these cases, the full time evolution gives a more complete description of the decay.}.
\begin{figure}[tb!]
    \centering
    \includegraphics[width=\columnwidth]{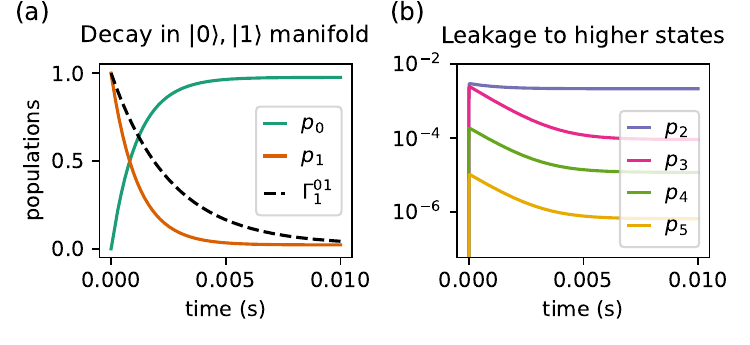}%
    \caption{Evolution of state population under transition matrix calculation, for dielectric loss at $\Phi_\text{ext}/\Phi_0 = 0.3$. 
    (a) Decay in the $\ket{0}, \ket{1}$ manifold for initialization in $\ket{1}$. $\Gamma_1^{01} = \Gamma^{0\rightarrow1} + \Gamma^{1\rightarrow 0}$ is the relaxation rate obtained from Fermi's golden rule (2-level model). (b) Leakage to higher states, showing negligible population transfer.}
    \label{supfig:TM-pops-decay}
\end{figure}
\begin{figure}[t!]
    \centering
    \includegraphics[width=0.95\columnwidth]{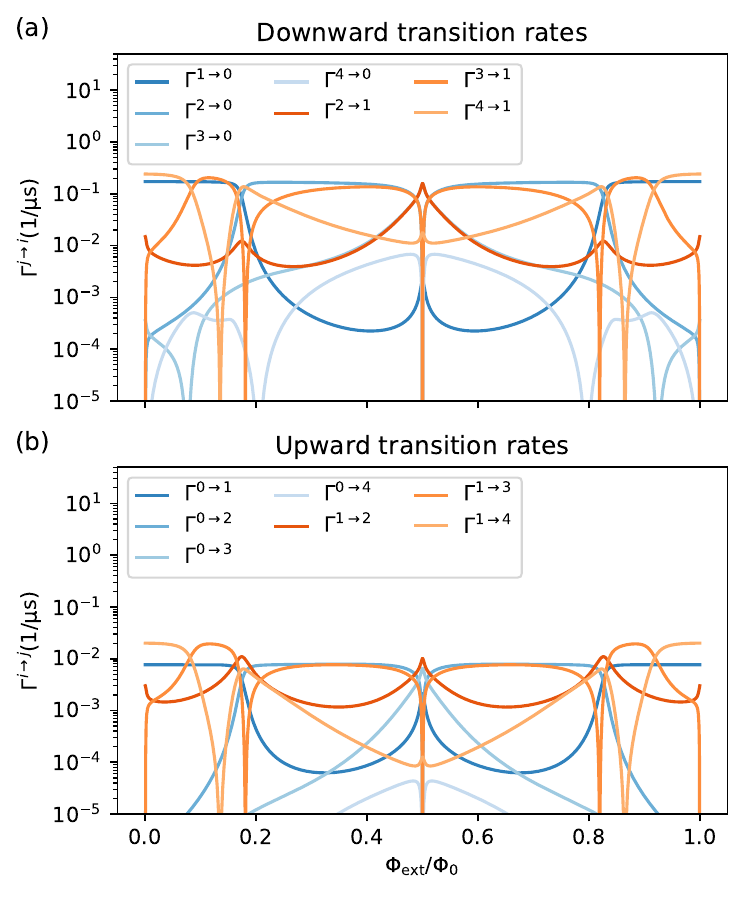}
    \caption{Transition rates for dielectric loss.}
    \label{supfig:diel-loss-trans-rates}
\end{figure}
As an example, we examine dielectric loss at $\Phi_\text{ext} / \Phi_0 = 0.3$. We compare the population evolution of the entire system in the $\{\ket{0}, \ket{1} \}$ subspace with the single decay rate obtained by truncating to two levels (see Fig.~\ref{supfig:TM-pops-decay}a). It is apparent that the lifetime is shorter when influenced by the higher levels. Inspecting the population buildup in the higher states shows that the leakage remains minimal (Fig.~\ref{supfig:TM-pops-decay}b).
We then extend this calculation across a full flux quantum (Fig.~\ref{supfig:N-level-model-compare}b), which shows that for dielectric loss, the two- and $N$-level model decay times display the largest difference at intermediate flux biases.
In contrast, both models are nearly identical at the half- and integer-flux bias points, indicating that the higher levels do not contribute appreciably there.

We gain further intuition by looking at the calculated transition rates versus flux bias for transitions up to $N = 4$ (Fig.~\ref{supfig:diel-loss-trans-rates}, higher transitions not shown for clarity). As can be seen at $\Phi_\text{ext} / \Phi_0 \approx 0.3$, the $\ket{1} \rightarrow \ket{2}$ excitation rate is comparable in magnitude to the $\ket{1} \rightarrow \ket{0}$ relaxation rate, making it likely that a qubit in the $\ket{1}$ state transitions to either $\ket{0}$ or $\ket{2}$, with the $\ket{2}$ state subsequently relaxing quickly back to $\ket{0}$. Including the non-computational levels accounts for multiple decay channels out of and back into the qubit manifold, increasing the effective depolarization rate as a result. 
At half-flux, the $\ket{0} \leftrightarrow \ket{1}$ transitions dominate as the frequency is far-detuned from the higher levels, while at zero-flux the charge matrix element $\matel{n}$ makes $\ket{1} \rightarrow \ket{0}$ relaxation the primary process. 
We find convergence of the rates for $N \geq 6$ levels.
Making the same comparison for the other loss mechanisms (see Fig.~\ref{supfig:N-level-model-compare}), we find that the contribution of the higher levels is most significant for dielectric loss in the range of the measured data.

\subsection{Multiple loss mechanisms}

To evaluate the population evolution under multiple loss mechanisms, we must evaluate a single rate matrix for the combination of models, namely
\begin{equation}
    B_\text{total} = \sum_{i \in \text{all models}} B_i,
\end{equation}
where $B_i$ is calculated from Eq.~\ref{eq:B-matrix-app} using the Fermi's golden rule expressions for the $i$th model. 
For the models used in Fig.~\ref{fig:t1-flux-baseline}a of the main text,
the decay rates of the eigenvector $\gamma_i$
are plotted in Fig.~\ref{supfig:Nlevel_full_model}a, along with the initial vector overlaps $|c_i|$ (Fig.~\ref{supfig:Nlevel_full_model}b).

We emphasize that the eigenvector rates of the combined noise model are not equal to the sum of the rates of the individual models, $\Gamma_1^\text{eff,tot} \neq \sum_{i\in \text{models}} \Gamma_{1i}^\text{eff}$, so features may appear in the full result that are not apparent from looking at a single loss mechanism alone. For example, resonant-like dips in $\Gamma_1^\text{eff}$ in Fig.~\ref{fig:t1-flux-baseline} appear near resonator crossings with higher fluxonium transitions.
% from enhanced Purcell-induced heating near resonator crossings with higher fluxonium transitions. 
For intuition, at $|\delta\Phi_\text{ext}| \approx 0.06$, where the resonator coincides with the fluxonium $\ket{0}\leftrightarrow\ket{3}$ transition, the enhanced Purcell-induced heating $\Gamma_{0\rightarrow 3}^\text{Purcell}$ combined with the quick relaxation from dielectric loss $\Gamma_{3\rightarrow 1}^\text{diel}$ results in a drop in the apparent $\ket{0}\leftrightarrow\ket{1}$ transition time.
% and the fast Purcell-induced heating $\Gamma_{0\rightarrow 3}^\text{Purcell}$ combined with the quick relaxation from dielectric-loss $\Gamma_{3\rightarrow 1}^\text{diel}$ results in a drop in the apparent $\ket{0}\rightarrow\ket{1}$ transition time. 

\subsection{Evaluating exponential behavior}

\begin{figure}[tb!]
    \centering
    \includegraphics[width=\columnwidth]{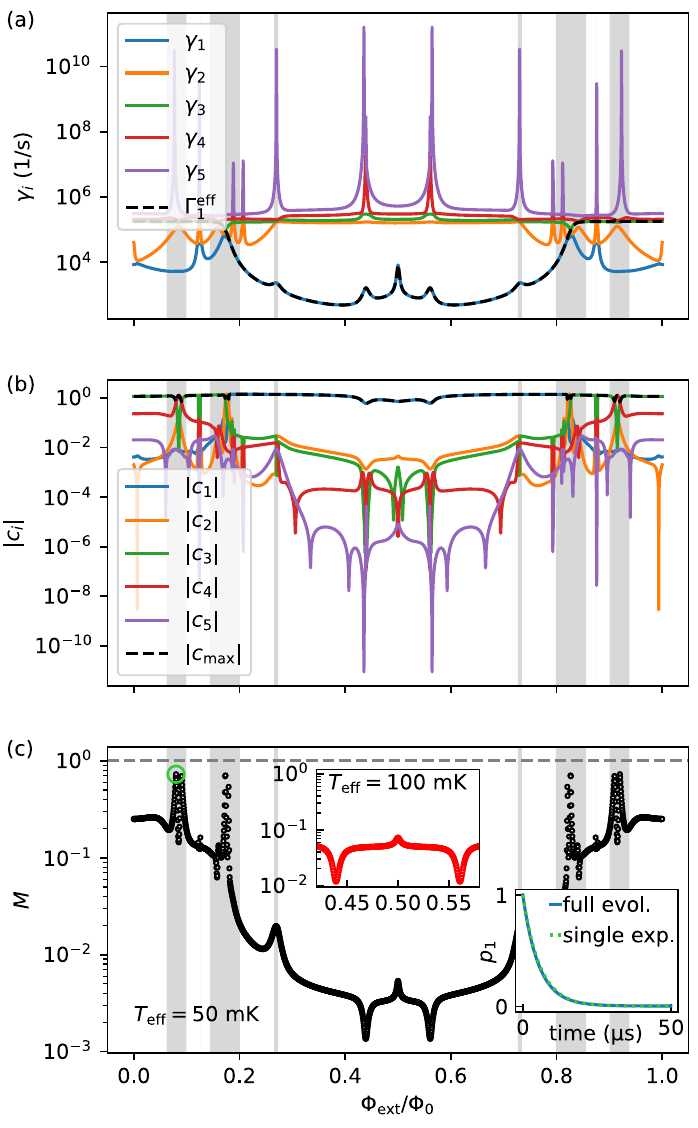}
    \caption{Eigenvectors of the $N$-level system. (a) Decay rates of the non-steady-state eigenvectors of the $N$-level system under a combined model of $1/f$ flux noise, dielectric loss, Purcell decay, and QP tunneling (same parameters as Fig.~\ref{fig:t1-flux-baseline}). Black dashed line indicates the eigenvalue representing the dominant decay vector in the $\{ \ket{0}, \ket{1} \}$ subspace, denoted $\Gamma_1^\text{eff}$. (b) Eigenvector coefficients $|c_i|$ in Eq.~\ref{eq:p-t-eigenvector-evolution} given the system being initialized into the first excited state ($p_{1} = 1, p_i = 0$ for $i\neq1$). Black dashed line follows the maximum mode overlap at each flux bias (with the corresponding eigenvalue plotted in part (a)). (c) Population of initial vector in non-dominant eigenvectors $M$ (Eq.~\ref{eq:M-metric}). Main: $T_\text{eff} = 50~\mathrm{mK}$, $T_\text{res} = 70~\mathrm{mK}$. Top inset: $T_\text{eff} = T_\text{res} =  100~\mathrm{mK}$, with flux range corresponding to where we measured temperature dependence. Bottom inset: solid line shows simulated evolution of $p_1(t)$ at flux bias corresponding to largest value of $M$ (i.e., least exponential decay, green circled point in main panel). Dotted line shows decay of a single exponential component with the eigenvalue of the dominant eigenvector. Comparison of the two traces reveals that full numerical solution is largely indistinguishable from single exponential decay. }
    \label{supfig:Nlevel_full_model}
\end{figure}

Although in general the solution $\vec{p}(t)$ contains multiple exponentially decaying terms, the resulting dynamics can often be approximately characterized by a single rate if there is a dominant exponentially decaying eigenmode in the initial state, and the overlap with the remaining decay modes is small.
We quantify the validity of the approximation by calculating the difference vector $\vec{\delta}= \vec{p}_0 - c_0\vec{v}_0 - c_{k_\text{max}} \vec{v}_{k_\text{max}}$, which represents the initial probability contained in the other (non-dominant) eigenmodes. 
We note that $\vec{\delta}$ is independent of the choice of normalization of the eigenvectors.  
The length of this difference vector $\vec{\delta}$ then serves as a metric to characterize how exponential the resulting dynamics are,
\begin{equation}
    M \equiv  ||\vec{\delta}|| = \sqrt{\sum_j \delta_j^2}, \label{eq:M-metric}
\end{equation}
\noindent with smaller values of $M$ corresponding to more exponential behavior. 
We plot $M$ as a function of flux at $T_\text{eff} = 50~\mathrm{mK}$ (Fig.~\ref{supfig:Nlevel_full_model}c) and $T_\text{eff} = 100~\mathrm{mK}$ (Fig.~\ref{supfig:Nlevel_full_model}c, top inset) for the flux-bias ranges that we consider, finding that the overlap with the other decay modes remains under $\sim 10 \%$ at most flux biases.
Even at the bias with the largest value of $M$, the full evolution of $p_1$ is qualitatively well-described by a single exponential decay (bottom inset of Fig.~\ref{supfig:Nlevel_full_model}c).

Near regions where avoided level crossings occur in the system, the predicted behavior is less exponential, and a single $\Gamma_1^\text{eff}$ becomes insufficient to describe the dynamics. We find that these regions are correlated with larger leakage state populations and a dependence on the choice of initial state ($\ket{0}$ or $\ket{1}$). Therefore, we apply this model with the caveat that the results may be less reliable in regions of strong hybridization. We leave an experimental verification of the multi-level decoherence and highly non-exponential dynamics to future work.

\subsection{Effect of leakage state mis-assignment on measured $T_1$}
\begin{figure}[tb!]
    \centering
    \includegraphics[width=\columnwidth]{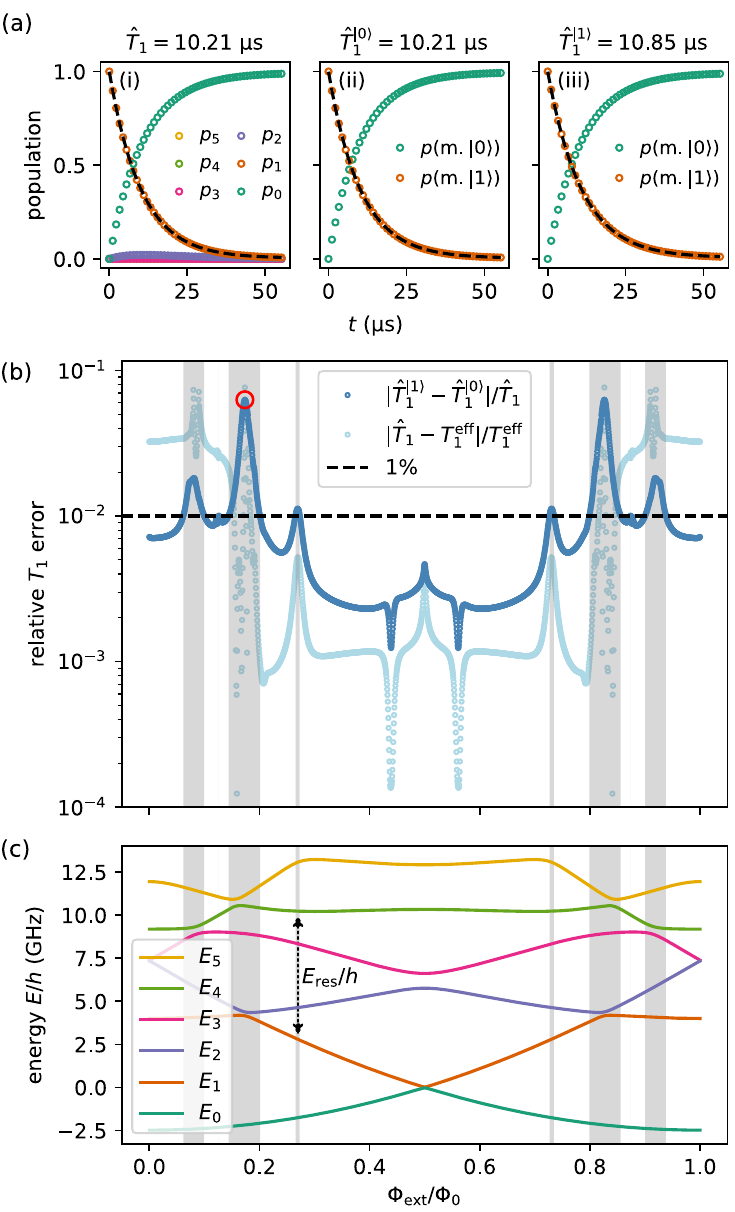}
    \caption{Simulated effect of leakage state mis-assignment. (a) Simulated populations and their fits at $\Phi_\text{ext} / \Phi_0 = 0.174$, corresponding to the largest relative error $| \hat{T}_1^{\ket{1}} - \hat{T}_1^{\ket{0}}|/\hat{T}_1$. Simulations of $T_1$ with (a.i) no assignment error, (a.ii) assignment of leakage states to $\ket{0}$, and (a.iii) assignment of leakage states to $\ket{1}$. Dashed line indicates fit to exponential decay. (b) Flux dependence of relative error from leakage state mis-assignment (dark blue) and difference between time constant obtained from fit and from eigenvalue (light blue), see text. Gray shaded regions correspond to flux biases where $| \hat{T}_1^{\ket{1}} - \hat{T}_1^{\ket{0}}|/\hat{T}_1 > 1\%$. Red circle indicates the ``worst case'' error for the example shown in panels (a.i - a.iii). (c) Qubit energy spectrum, obtained theoretically from fitted Hamiltonian parameters, plotted with same shading as panel (b) to show that regions of large error correspond to flux biases around avoided level crossings. The band around $\Phi_\text{ext} = 0.27\Phi_0$ aligns with the readout resonator crossing the qubit $\ket{1} \leftrightarrow \ket{4}$ transition.} 
    \label{supfig:leakage-readout-error}
\end{figure}

Here we address the possibility that, under the $N$-level decoherence model, leakage states ($\ket{n}$ with $n>1$) may be populated but not distinguished in qubit readout. Instead, they may be mis-assigned to $\ket{0}$ or $\ket{1}$. We consider the theoretical effect of this leakage state mis-assignment on the extracted $T_1$ in the limit of small leakage.  

We consider a readout scheme where single-shot measurement events are classified into $\ket{0}$ or $\ket{1}$ by which distance to the corresponding cluster mean in IQ space is smaller. There are two extreme scenarios of mistaken assignment: one in which all leakage states are assigned to $\ket{0}$, and one in which they are assigned to $\ket{1}$. Let $p(\text{measure}\ket{x})$ be the probability of measuring state $\ket{x}$. Depending on which mistaken assignment is performed, these are given by the following probabilities:
\begin{align}
    p(\text{measure} \ket{0}) &= p_0 (t)+\sum_{n\geq2} p_{n}(t) \\
    p(\text{measure} \ket{1}) &= p_{1}(t)
\end{align}
for mis-assigning to $\ket{0}$, and 
\begin{align}
    p(\text{measure} \ket{0}) &= p_0 (t)  \\
    p(\text{measure} \ket{1}) &= p_{1}(t) +\sum_{n\geq2} p_{n}(t)
\end{align}
for mis-assigning to $\ket{1}$, where the $p_i$ denote the populations calculated from the $N$-level decoherence model. To quantify the effect on the extracted $T_1$ value, we define the following: 
Let $\hat{T}_1$ be the estimator of $T_1$ obtained by fitting the ``true'' $p_1 (t)$ to an exponential decay profile, with no assignment error (see Fig.~\ref{supfig:leakage-readout-error}a.i).
Further, let $\hat{T}_1^{\ket{x}}$ be the estimator obtained by fitting the ``measured'' excited state trace $p(\text{measure} \ket{1})$ to an exponential decay profile, given that the leakage states are mistakenly assigned to $\ket{x}$ (see Figs.~\ref{supfig:leakage-readout-error}a.ii-iii). This simulates the analysis of a standard $T_1$ measurement in the absence of sampling noise. 
These quantities are both distinct from $T_1^\text{eff} \equiv 1/\Gamma_1^\text{eff}$, the decay time of the dominant eigenvector of the system, discussed above. In Fig.~\ref{supfig:leakage-readout-error}b we plot $| \hat{T}_1^{\ket{1}} - \hat{T}_1^{\ket{0}}|/\hat{T}_1$, the relative error of the estimator of $T_1$ due to leakage state mis-assignment, and $| \hat{T}_1 - T_1^\text{eff}|/T_1^\text{eff}$, the relative difference between the estimator of $T_1$ and the effective decay time. 
For our parameters, both errors are at the percent level of the $T_1$ value (effective or estimator), and at many flux biases below 1\%. The first error tells us that leakage state mis-assignment makes only a small difference to the extracted $T_1$ value, and the second shows that the result of the eigenvalue analysis method is close to that obtained by the estimator. As shown in Fig.~\ref{supfig:leakage-readout-error}c, the larger errors of approximately $1-7\%$ occur near avoided level crossings, where thermal excitation is enhanced across the small energy gap.
We note that, when the readout signal is taken to be the average value of $S_{21}$ rather than the result of a discrete state assignment, the error in $T_1$ is similarly at the percent level or less for all flux biases.

\section{Potential TLS sources of $T^d$ scaling} \label{app:TLS}
In our data we observe an apparent $T^{\beta_2}$ scaling of the charge noise, with $\beta_2 \approx 3$ if we neglect the contributions from non-computational levels. 
In the main text we considered two ways in which a power-law temperature dependence could arise from the standard tunneling model (STM).
Here we provide some detail on the expressions for resonant- and relaxation-type losses resulting from coupling to defect ensembles. We also offer some discussion on further topics of exploration.  

\subsection{Resonant absorption}
The contribution to the dielectric loss tangent from resonant TLS absorption is~\cite{Phillips1987TwoLevelStates, Behunin2016Dimensional}
\begin{align}
\begin{split}
    \tan\delta_\text{res}(\omega, \bar{n}, T) &= \frac{\pi P|\mathbf{p}|^2}{3\epsilon_0\epsilon_r} \frac{ \tanh\left( \frac{\hbar\omega}{2k_B T} \right)}{\sqrt{1+\bar{n}/n_c}} %\\
    % & \equiv \tan\delta_{0,\text{res}} \frac{ \tanh\left( \frac{\hbar\omega}{2k_B T} \right)}{\sqrt{1+\bar{n}/n_c}},
\end{split}\label{eq:tls_res_abs}
\end{align}
where $\mathbf{p}$ is the electric dipole moment of the TLS, $\epsilon_r$ is the material relative permittivity, $\bar{n}$ is the average circulating photon number, and
$P$ characterizes the defect density of states $f(\Delta, \Delta_0) = P/\Delta_0$, where a uniform distribution of TLS asymmetry energies $\Delta$ and log-uniform distribution of tunneling energies $\Delta_0$ is assumed.
The critical photon number $n_c$ is related to the TLS coherence times $n_c \propto (\tau_1\tau_2)^{-1}$, so for large photon numbers the loss scales approximately as $\tan\delta_\text{res} \propto (\tau_1\tau_2)^{-1/2} \tanh\left( \frac{\hbar\omega}{2k_B T} \right)$. A TLS lifetime scaling as $\tau_1^{-1} \propto T^\beta$ with $\tau_2 = 2\tau_1$ may result in a power-law dependence in the qubit depolarization rate $\Gamma_1^\text{TLS, res} \propto T^\beta |\langle 0|\hat{n}|1 \rangle |^2$.

\subsection{Relaxation absorption}
To calculate the relaxation contribution, one integrates the individual TLS contribution over the defect density of states (DOS). Following Refs.~\cite{Phillips1987TwoLevelStates} and ~\cite{Behunin2016Dimensional}, similar to Ref.~\cite{Mittal2024Annealing}, this gives for the relaxation contribution to the loss tangent,
\begin{align}
\begin{split}
    \tan&\delta_\text{TLS,rel}(\omega,T)  = \frac{P|\mathbf{p}|^2 \omega}{6 \epsilon_0 \epsilon_r k_BT}  
     \int_0^\infty dE \int_{\tau_\text{1,min}(E)}^\infty  d\tau_1 \\
    & (1-\tau_\text{1,min}(E)/\tau_1)^{1/2}  \text{sech}^2 \left( \frac{E}{2k_BT} \right) \frac{1}{1 + \omega^2 \tau_1^2},
\end{split}
\end{align}
where a change of variables has been made to write the DOS in terms of TLS energy $E = \sqrt{\Delta^2 + \Delta_0^2}$ and relaxation time $\tau_1$: $g(E,\tau_1) dE d\tau_1 = f(\Delta, \Delta_0) \frac{\Delta_0 E}{2\Delta \tau_1} d\Delta d\Delta_0$.
For phonon-limited relaxation, the expression for $\tau_1$ is given by~\cite{Behunin2016Dimensional}
\begin{equation}
    \tau_1^{-1}(E) = \frac{\gamma^2}{v^{d+2}} \frac{\pi S_{d-1}}{(2\pi)^d} \frac{E^{d-2} \Delta_0^2}{\hbar^{d+1}\rho_d} \coth \left( \frac{E}{2k_BT} \right), \label{supeq:TLS-T1}
\end{equation}
where $\gamma$ is the (average) elastic dipole moment of the TLS, $v$ the speed of sound in the material, $\rho_d$ the $d$-dimensional material density, and $S_{d}$ the $d$-dimensional unit hypersphere surface area. 
$\tau_\text{min}$ is found by evaluating Eq.~\ref{supeq:TLS-T1} for $\Delta_0=E$.
Note that the $\text{sech}^2 \left( \frac{E}{2k_BT} \right)$ term in the integral is suppressed for defects with $E \gg 2k_BT$, making this process primarily relevant for high temperatures and/or low frequencies, relevant for the heavy fluxonium system.
Evaluating this integral in the high-frequency limit gives a power law dependence in temperature:
\begin{align}
\begin{split}
    \lim_{\omega\tau_\text{min} \gg 1} \tan\delta_\text{TLS,rel}(\omega,T) &= \frac{ P |\mathbf{p}|^2 }{9 \epsilon_0 \epsilon_r} \xi \mathcal{I}_d \frac{(2k_BT)^d}{\omega} %\\
    % &\equiv \frac{\tan\delta_{0,\text{rel}}}{\omega} \left( \frac{T}{T_0} \right)^d
\end{split}
\end{align}
where $\xi = \frac{\gamma^2}{v^{d+2}} \frac{\pi S_{d-1}}{(2\pi)^d} \frac{1}{\hbar^{d+1}\rho_d}$ and $\mathcal{I}_d = \int_0^\infty u^d \text{sech}^2(u) \coth(u) du$.
The corresponding qubit relaxation rate scales as $\Gamma_1^\text{TLS, rel} \propto  \omega^{-1} T^d \coth(\hbar\omega/2k_BT) |\langle 0|\hat{n}|1 \rangle |^2$, where $d$ may vary between 1 and 3 depending on the effective system dimensionality.

\subsection{Discussion}
As noted in the main text, 
while both resonant and relaxation absorption could plausibly yield the power-law temperature scaling observed in our data, neither mechanism offers a fully satisfying description. This is, of course, further complicated by the presence of higher levels in the system. 

We also acknowledge prior evidence from measurements of frequency noise in microwave resonators~\cite{burnettEvidenceInteractingTwolevel2014}, as well as the time dependence of $T_1$ fluctuations in transmons~\cite{Klimov2018Fluctuations}, suggesting better consistency with an interacting defect model beyond the STM~\cite{faoro2015interacting}. Such a model groups defects into two categories: high-frequency, coherent TLS and low-frequency, thermally active TLS. In most studies of superconducting resonators and qubits at $\sim5~\mathrm{GHz}$ and millikelvin temperatures, it is presumed that the TLS interacting directly with the circuit are predominantly in their ground state. Through dispersive coupling with low-frequency TLS undergoing thermal fluctuations, the high-frequency TLS may diffuse in frequency space, in turn causing fluctuations in qubit $T_1$. In low-frequency qubits, however, such a distinction may be difficult to make, as thermally fluctuating TLS may also interact resonantly with a qubit transition. Further probing this regime of TLS behavior, e.g., via time-dependent $T_1$ measurements, could offer valuable insight into the properties of the TLS baths affecting superconducting qubits.

\section{Dependence of qubit parameters on magnetic field} \label{app:B-field-qb-params}
\begin{figure}[tb!]
    \centering
    \includegraphics[width=\columnwidth]{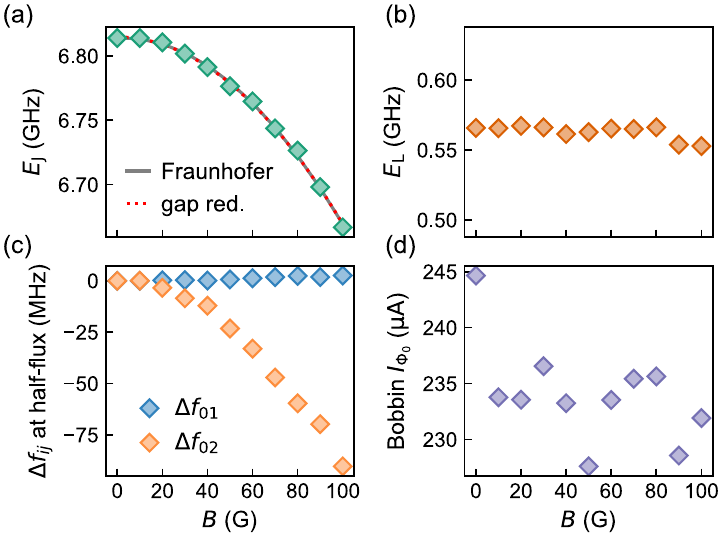}
    \caption{Dependence of qubit parameters on magnetic field. Fits were obtained holding $E_C=0.9544~\mathrm{GHz}$ fixed for each field. Half-flux frequency $f_{01}$ changes by only $5\%$ between 0 and 100 G. Fraunhofer: Fit to $E_J(B)=E_J(0) \left| \text{sinc}\left( \frac{B - B_\Delta}{B_{\Phi_0}} \right)\right|$~\cite{orlando1991foundations}.}
    \label{supfig:bfield-qb-params}
\end{figure}

Here we present measurements of the qubit Hamiltonian parameters as a function of applied in-plane field. We tracked the Hamiltonian parameters by performing a fit to two-tone spectroscopy data including transitions up to $\ket{0} \rightarrow \ket{3}$. At each field, we fixed the charging energy to the zero-field value ($E_C = 0.954~\mathrm{GHz}$) and fit the spectrum to obtain $E_J$ and $E_L$, plotted in Fig.~\ref{supfig:bfield-qb-params}(a) and (b), respectively. In Fig.~\ref{supfig:bfield-qb-params}(c) we plot the change in half-flux frequency of the $\ket{0}\rightarrow\ket{1}$ and $\ket{0} \rightarrow \ket{2}$ transitions with respect to the zero-field values, obtained by diagonalizing the Hamiltonian with the parameters extracted at each field. Fig.~\ref{supfig:bfield-qb-params}(d) shows the field dependence of the out-of-plane flux periodicity in terms of bias current, $I_{\Phi_0}$. 

Parallel flux interference in the junction creates a Fraunhofer-like modulation of the critical current $I_c \propto E_J$ with in-plane field~\cite{orlando1991foundations}. We fit the measured data to $E_J(B) = E_J^0 \left| \text{sinc} \left( \frac{B - B_\Delta}{B_{\Phi_0}} \right) \right|$, where $B_{\Phi_0} = 857~\mathrm{G}$ is the periodicity and $B_\Delta = 2.2~\mathrm{G}$ is an offset field. 
The extracted periodicity for one flux quantum $\Phi_0 = B_{\Phi_0}A$ gives an estimate of the effective junction cross section area $A = ld$, where $l$ is the junction length and $d = 2\lambda + a$, with $\lambda \approx 50~\mathrm{nm}$ the Al penetration depth, and $a\approx 1~\mathrm{nm}$ the oxide thickness.
From these values we estimate the junction length to be $l \approx 240~\mathrm{nm}$, in good agreement with the design target of $\approx 230~\mathrm{nm}$.
This is consistent with previous works which found that the $E_J$ variation in transmons with in-plane field was in good agreement with the Fraunhofer modulation from flux penetration~\cite{schneiderTransmonQubitMagnetic2019, Krause2022Magnetic, Krause2024Quasiparticle}.

We can alternatively consider the decrease in $E_J$ in terms of a reduction of the superconducting gap with field from Ginzburg-Landau theory: $E_J(B) \propto \Delta(B) = \Delta_0 \sqrt{1 - B^2/B_c^2}$, where in the limit of $T \ll T_c$, $I_c \propto \Delta$~\cite{tinkham2004introduction}. 
While the two models are effectively indistinguishable in the range of $B$ we measured, the extracted estimate for the critical field $B_c = 487~\mathrm{G}$ is not consistent with a complete suppression of critical current at $B = B_c$.
Even if this were the case, the excess QP density resulting from a gap suppression of only $\sim 2 \%$ is of order $\Delta x_\text{qp} \approx 1 \times 10^{-22}$ at 50 mK, so the magnetic field dependence of $T_1$ is unlikely to be explained by excess thermal quasiparticle generation from a reduction of the gap. 

We note that the qubit frequency $f_{01}$, phase matrix element $\matel{\phi}$, and charge matrix element $\matel{n}$ all change by $< 8\%$ between $B = 0~\mathrm{G}$ and $B = 100~\mathrm{G}$ within the flux range considered for the magnetic field dependence. At half-flux, the changes in $f_{01}$ and $\matel{n}$ are both $\sim5\%$, while the change in $\matel{\phi}$ is only $\sim 0.04\%$. Since these variations are small, we choose to ignore their effect on the conversion from $T_1$ data to noise spectra, and assume the same qubit parameters at all fields in the analysis.

\section{Supplemental Magnetic field data} \label{app:more-B-field-data}
% Gamma1 contributions and T1 box plot
\begin{figure}[tb!]
    \centering
    \includegraphics[width=0.95\columnwidth]{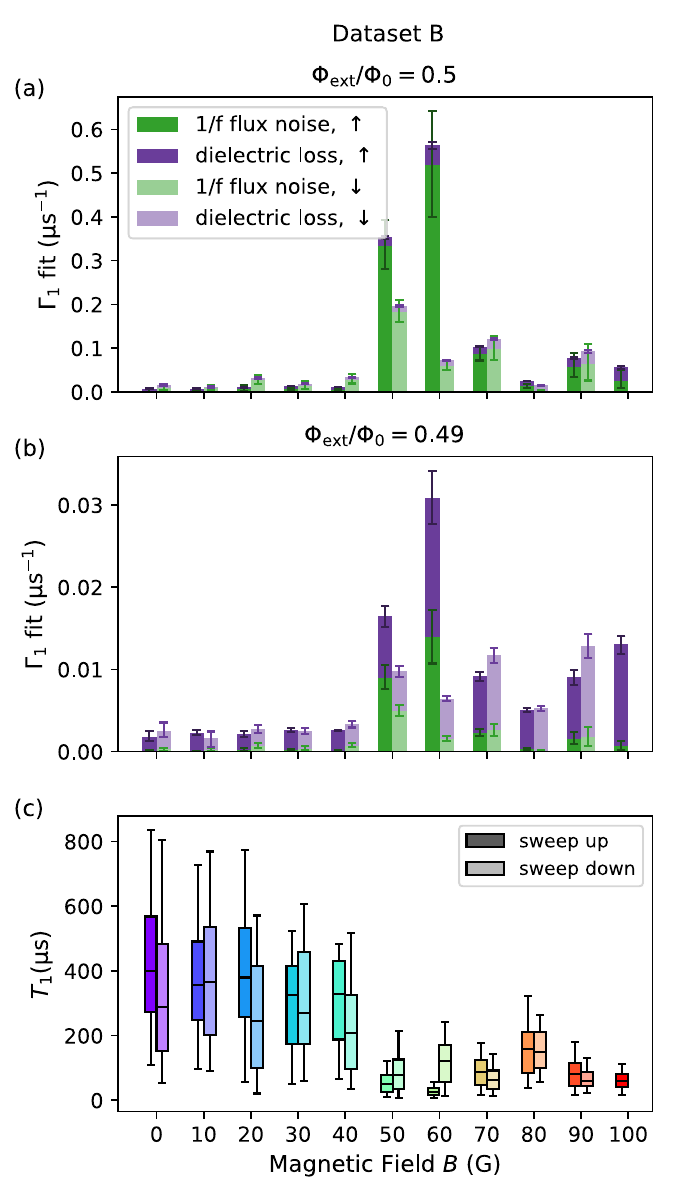}
    \caption{Magnetic-field dependence for dataset B.}
    \label{supfig:B-summary-dataset1B}
\end{figure}
% 
% Spectrum parameter summary plots
\begin{figure}[tb!]
    \centering
    \includegraphics[width=\columnwidth]{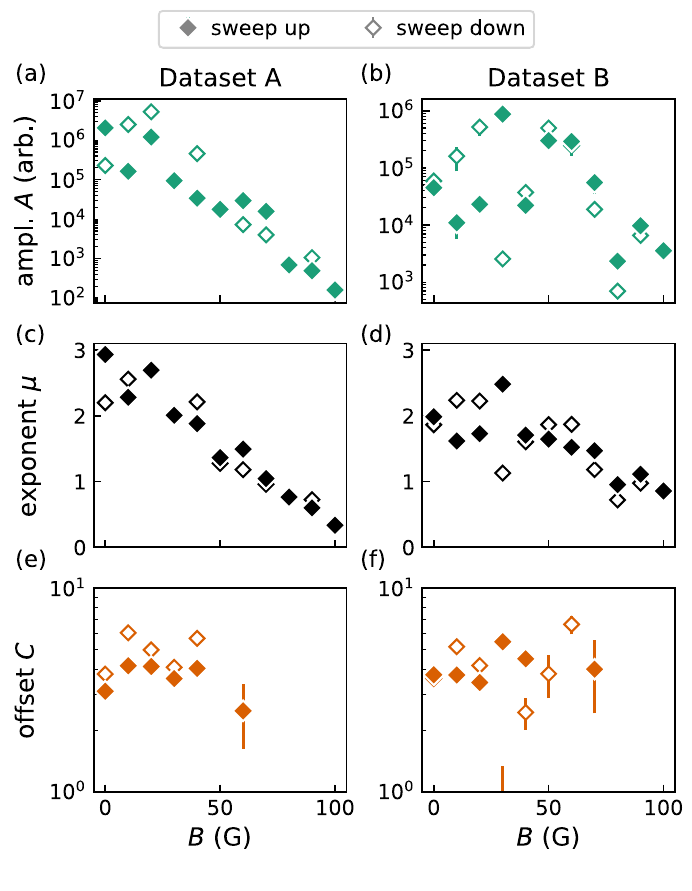}
    \caption{Fit parameters for both magnetic field dependence datasets. For each field, $\Gamma_1/\matel{n}^2$ data were fit to to $y = A/f_{01}^\mu + C$. (a-b) Amplitude $A$, (c-d) exponent $\mu$, (e-f) offset $C$. At some field values, the relative parameter uncertainty in the white noise offset was large ($\sigma_C / C > 1$); for these fields, data were re-fit with the constraint $C = 0$ and the offset was not included in the figure. Error bars correspond to one standard deviation of the fit parameters. }
    \label{supfig:B-spectrum-params-both-datasets}
\end{figure}

\begin{figure*}[h!]
    \centering
    \includegraphics[width=\textwidth]{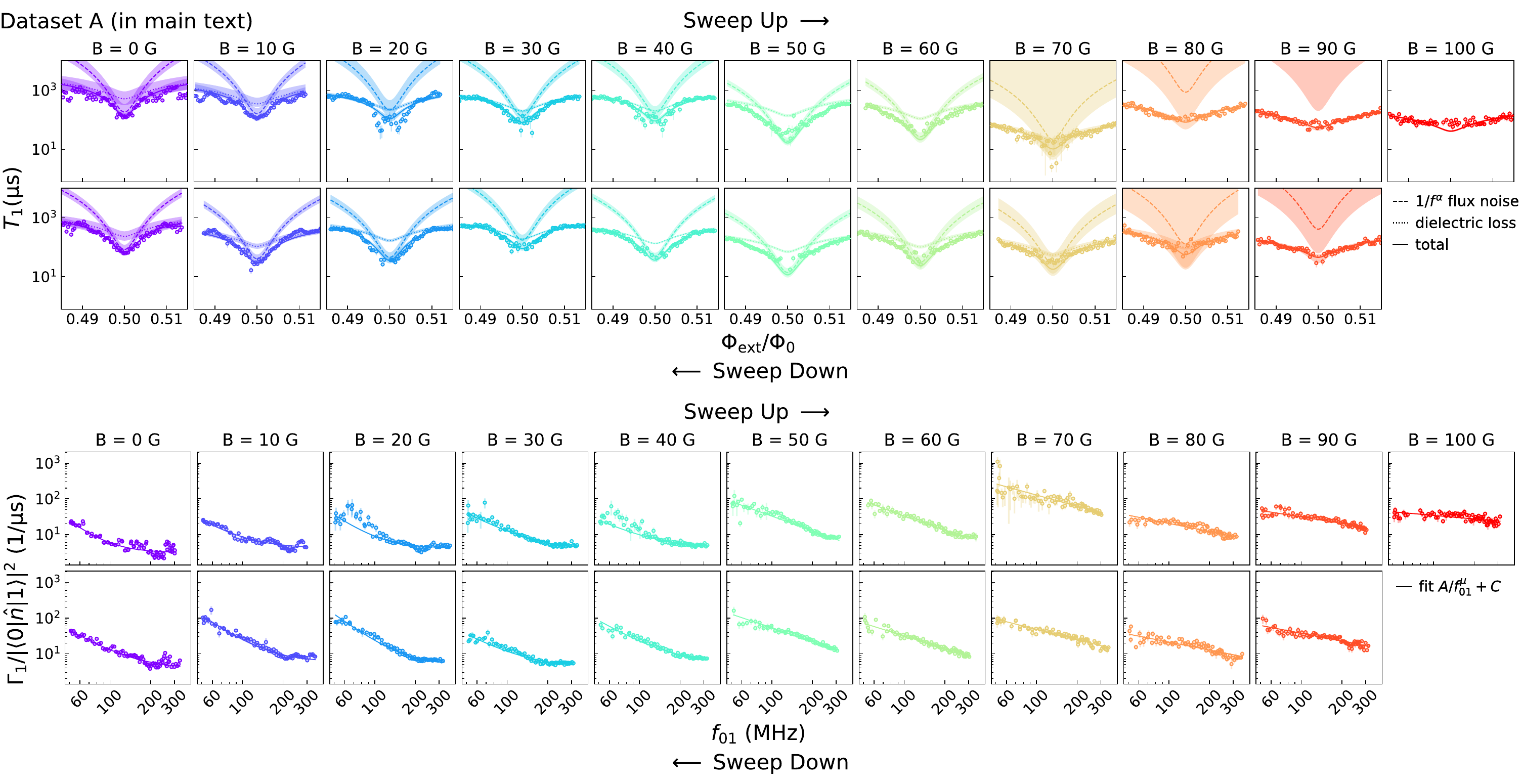}
    \caption{Full $B$-field dataset A, in main text.}
    \label{supfig:B-T1-dataset2}
\end{figure*}

\begin{figure*}[h!]
    \centering
    \includegraphics[width=\textwidth]{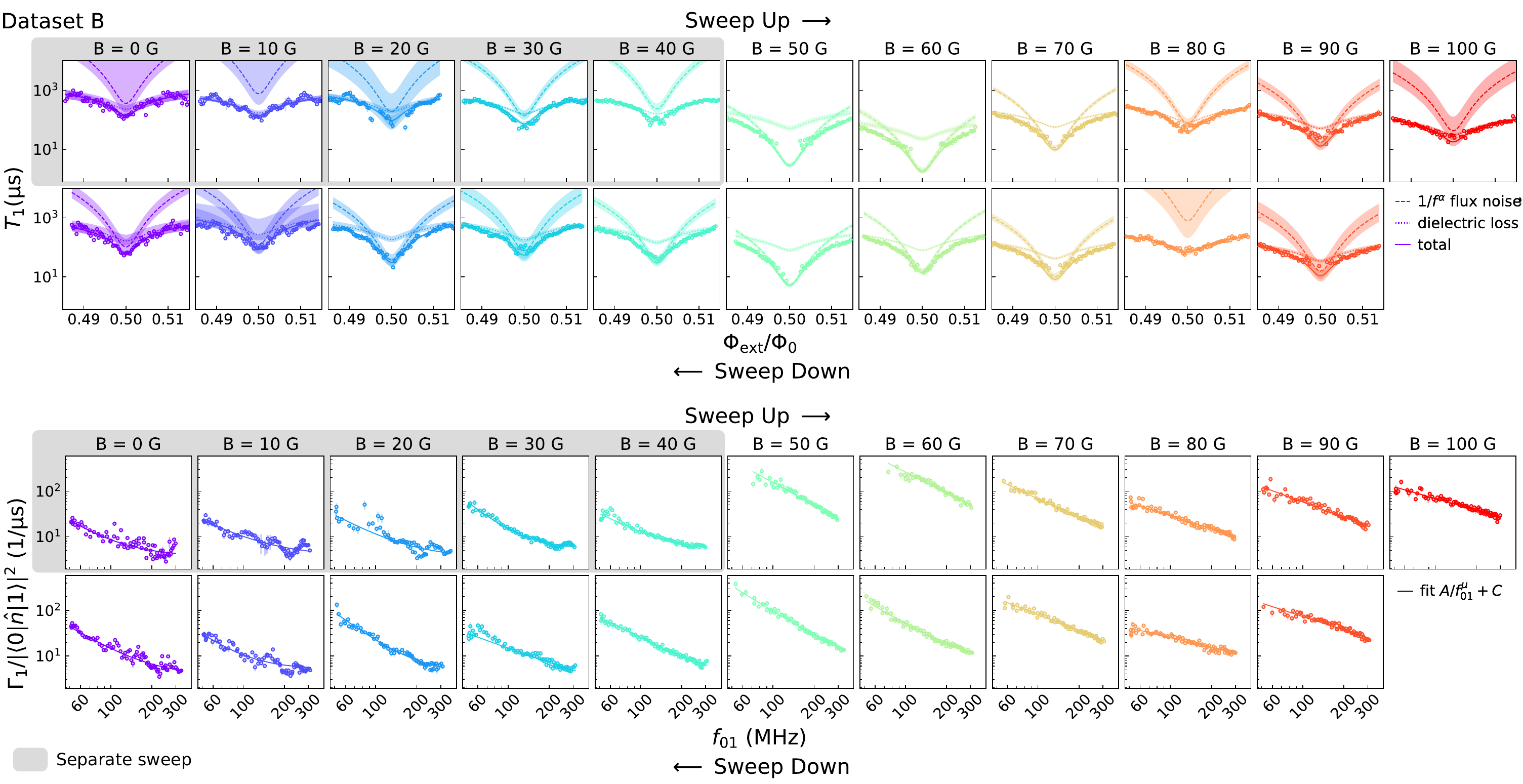}
    \caption{Full $B$-field dataset B. Note that here, the upwards sweep to 40 G was taken in a different sweep than 50 G $\rightarrow$ 100 G $\rightarrow$ 0 G.}
    \label{supfig:B-T1-dataset1}
\end{figure*}

Here we present supplemental data and analysis for two magnetic field sweeps conducted for the experiment.
In the following, Dataset A refers to the measurements shown in the main text, while Dataset B corresponds to an additional field sweep, taken in a different cooldown.
Fig.~\ref{supfig:B-spectrum-params-both-datasets} shows the full set of estimated parameters obtained by fitting $\Gamma_1/\matel{n}^2$ versus frequency to $y = A/f_{01}^\mu + C$. 
We note that, at fields where the relative fit uncertainty in the white-noise level $\sigma_C/C > 1$, we refit the data holding $C$ fixed to zero, hence the missing points in Fig.~\ref{supfig:B-spectrum-params-both-datasets}e-f.
Fig.~\ref{supfig:B-summary-dataset1B} plots the estimated contributions of $1/f^\alpha$ flux noise and dielectric loss as well as the box plot of $T_1$ values as a function of field, for Dataset B.
The top (bottom) of Figs.~\ref{supfig:B-T1-dataset2} and~\ref{supfig:B-T1-dataset1} show the full set of $T_1$ versus flux ($\Gamma_1/\matel{n}^2$ versus frequency) measurements for both datasets, along with their respective fits. We note that Dataset B was taken in two separate stages, but within the same cooldown: 0 G $\rightarrow$ 50 G, and 50 G $\rightarrow$ 100 G $\rightarrow$ 0 G.
We compare the two datasets by applying the same fitting procedures as described in the main text.
For the $T_1$ fits, we set $\alpha = 0.62$ and $\epsilon = 0.31$ for $S_{\Phi}$ and $S_{Q}$, respectively, and fit $T_1$ versus flux to $1/T_1 = \Gamma_1^\Phi + \Gamma_1^{Q}$ at each magnetic field (Figs.~\ref{supfig:B-T1-dataset2} and~\ref{supfig:B-T1-dataset1}).
The estimated fit contributions from flux noise and dielectric loss at the two selected flux biases are plotted in Figs.~\ref{supfig:B-summary-dataset1B}a-b for Dataset B.
Comparing to Figs.~\ref{fig:T1-B-sweep}b-c in the main text, a qualitatively similar trend is observed: an overall increase in dielectric loss, along with a sharp rise in flux noise at $50~\mathrm{G} \lesssim B \lesssim 80~\mathrm{G}$. 
The measured $T_1$ distributions around half-flux also show a qualitatively similar non-monotonically decreasing trend with field (Fig.~\ref{supfig:B-spectrum-params-both-datasets}c). 
A similar comparison can be made for the normalized rates. Fig.~\ref{supfig:B-spectrum-params-both-datasets} shows the full set of fit parameters for both datasets to the function $A/f_{01}^\mu + C$. 
The exponent $\mu$ shows a roughly consistent decreasing tend in both, while the amplitude $A$ exhibits more variation in Dataset B. 
We attribute these quantitative differences to the possibility of different trapped background fields between cooldowns, potential rearrangement of defects, and/or hysteresis in the response to different local field variations across the device. Nonetheless, the consistency in qualitative trends suggests that the underlying mechanisms show a response to magnetic fields.

\subsection{Independence of B-field trends on choice of fixed parameters and inclusion of heating effects}

Throughout the analysis of the magnetic field dependence, we made two key assumptions which we address here. 
First, we held $\alpha$ and $\epsilon$ fixed in our fits in an effort to reduce the chances of over-fitting our $T_1$ data.
These parameters characterize the noise exponent of the low-frequency flux and charge noise spectra ($\alpha$ and $1-\epsilon$, respectively). 
Of course, these are not generally constant with applied field, and elucidating the trends more carefully may shed light on the relevance of mechanisms such as spin diffusion, which
relates regions of different frequency scalings to characteristic length scales in the system~\cite{lantingEvidenceTemperaturedependentSpin2014}.
In our experiment, the tuning of the matrix elements and resulting sensitivity to both flux and charge noise made it impossible to disentangle the scaling of these two spectra.
However, to show that our conclusions were independent of the choice of these fixed parameters, we confirmed that repeating the analysis while varying $\alpha$ between 0.5 and 1.5 and $\epsilon$ between 0 and 0.7 yielded the same qualitative trends. 

Second, we used the two-level loss model for simplicity, excluding heating effects to non-computational states.
The effect of including the heating is to ``flatten'' the dielectric loss contribution with flux, and the result is that the half-flux region becomes more biased towards flux noise. 
While the estimates for the relative contributions of flux and charge noise change, the overall conclusion that dielectric loss increases with field remains consistent.
% 
% Full dataset A
% 
% Full dataset B
\clearpage
% 

% \FloatBarrier

% ---------------------------------------------------------------------------------------------------------
% References
% ---------------------------------------------------------------------------------------------------------
\bibliography{references}

% ---------------------------------------------------------------------------------------------------------
\twocolumngrid

\end{document}